\documentclass[12pt]{article}

\usepackage{amsmath}
\usepackage{amssymb}
\usepackage{amsfonts}
\usepackage{latexsym}
\usepackage{color}
\usepackage{soul}
\usepackage{tikz}
\usetikzlibrary{arrows,decorations.markings,calc}
\usetikzlibrary{decorations.pathmorphing,patterns}
\usepackage{graphicx}
\usepackage{subfigure}
\usepackage{color,epstopdf}
\catcode `\@=11 \@addtoreset{equation}{section}

\catcode `\@=12

\DeclareFontFamily{U}{skulls}{}
\DeclareFontShape{U}{skulls}{m}{n}{ <-> skull }{}
\newcommand{\skull}{\text{\usefont{U}{skulls}{m}{n}\symbol{'101}}}

\newcommand{\be}{\begin{equation}}
\newcommand{\en}{\end{equation}}
\newcommand{\bea}{\begin{eqnarray}}
\newcommand{\ena}{\end{eqnarray}}
\newcommand{\beano}{\begin{eqnarray*}}
\newcommand{\enano}{\end{eqnarray*}}
\newcommand{\bee}{\begin{enumerate}}
\newcommand{\ene}{\end{enumerate}}

\newcommand{\red}{\color{red}}

\newcommand{\A}{{\mathfrak A}}
\newcommand{\B}{{\mathfrak B}}

\newcommand{\Ac}{{\mathcal A}}
\newcommand{\Bc}{{\mathcal B}}
\newcommand{\Kc}{{\mathcal K}}

\newcommand{\mc}{\mathcal}

\newcommand{\D}{{\mc D}}
\newcommand{\V}{{\mc V}}

\newcommand{\E}{{\cal E}}
\newcommand{\F}{{\cal F}}

\newcommand{\Lc}{{\cal L}}

\newcommand{\kt}{\rangle}
\newcommand{\br}{\langle}

\newcommand{\1}{1 \!\! 1}

\newcommand{\Hil}{\mc H}

\begin{document}
\vspace*{-2cm}
\vspace{1cm}
\noindent This is the pre-peer reviewed version of the article submitted at Annals of Physics \\
\thispagestyle{empty}

\begin{center} {\Large \bf Coherent States of Graphene Layer with and without a $\mathcal{P}\mathcal{T}$-symmetric Chemical Potential}   \vspace{1cm}\\

{\large F. Bagarello}\\
  Dipartimento di Ingegneria,
Universit\`a di Palermo, 90128  Palermo, Italy\\
and I.N.F.N., Sezione di Catania, 95123 Catania, Italy\\
e-mail: fabio.bagarello@unipa.it\\

\vspace{2mm}

{\large F. Gargano}\\
Dipartimento di Ingegneria,
Universit\`a di Palermo, 90128  Palermo, Italy\\
e-mail: francesco.gargano@unipa.it\\

\vspace{2mm}

{\large L. Saluto}\\
Dipartimento di Ingegneria,
Universit\`a di Palermo, 90128  Palermo, Italy\\
e-mail: lidia.saluto@unipa.it\\

\end{center}

\vspace{0.5cm}

\begin{abstract}
\noindent
In this paper we construct different classes of coherent and bicoherent states for the graphene tight-binding model in presence of a magnetic
field, and for a deformed version where we include a $\mathcal{P}\mathcal{T}$-symmetric chemical potential $V$. 
In particular, the problems caused by the absence of a suitable ground state for the system is taken into account in the construction of these states, for $V=0$ and for $V\neq0$. We introduce ladder operators which work well in our context, and we show, in particular,  that there exists a choice of these operators which produce a factorization of the Hamiltonian. The role of broken and unbroken $\mathcal{P}\mathcal{T}$-symmetry is discussed, in connection with the strength of $V$.
\end{abstract}




\newpage

\section{Introduction}

Graphene's emergence as a two-dimensional material has revolutionized various scientific disciplines, due to its remarkable electronic, mechanical, and thermal properties. Since many years a lot of researchers started to discuss the physical aspects of graphene, and its concrete many applications. But the mathematical settings related to its description  also started to attract more mathematically oriented people, in view of the several interesting aspects, connected to it, which arise when dealing with the analytic aspects of graphene. We refer to \cite{Novoselov-2004,Zhang-2005,Novoselov-2005,Novoselov-2007,Geim-2009,CastroNeto-2009,DasSarma-2011}, and to \cite{graphenebook}, for a very partial list of contributions on the topic.  One of the most interesting features emerges particularly when a magnetic field is applied to it \cite{Zhang-2005,Novoselov-2005,Gusynin-2005}.

In \cite{BagaHatano_2016} a chemical potential was introduced in the system,  motivated by the possibility of simulating the effects of external fields and doping the graphene's electronic structure. This approach forced the authors in \cite{BagaHatano_2016} to construct a rich (non-Hermitian) quantum mechanical settings which includes the construction of suitable biorthogonal families of  eigenstates, which are different depending on the strength of the potential.

One of the features we will meet in the following is a $\mathcal{P}\mathcal{T}$ symmetry breaking
transition, \cite{benbook}. As widely known, this transition plays a pivotal role in non-Hermitian quantum mechanics, usually linked to systems that exhibit balanced gain and loss. In graphene, $\mathcal{P}\mathcal{T}$ symmetry and its breaking lead to the emergence of exceptional points—critical junctures where the properties of the system undergo some changes.

Central to the study we discuss here are, again, the families of biorthonormal states, whose definition is strongly determined by the width of the broken region caused from the chemical potential applied to graphene. By constructing specific annihilation operators in terms of these states, we propose a detailed analysis of bicoherent states,  somewhat inspired by some recent papers, \cite{Diaz-Fernandez-2017,Casti2020,Diaz-Negro-2021,Fernandez-Campa-2022,bagaop2024}. More explicitly, in our construction a special role will be played by the biorthonormal families of eigenvectors of the deformed Hamiltonian, and of its adjoint, describing graphene in one Dirac point in presence of chemical potential.

The paper is organized as follows. In  Section \ref{sect2} we consider the {\em standard situation}, i.e. graphene in absence of chemical potential $V$, and coherent states associated to it. This is already a non trivial task, since as it was already recently observed in \cite{bagaop2024}, it is not possible to talk of a ground state of the Hamiltonian, since the set of its eigenvalues is not bounded from above and from below. This implies that there is no {\em naturally chosen} eigenvector of the Hamiltonian which is annihilated by some relevant ladder operator. What we do here is to propose a possible construction which is linked to the one proposed in \cite{bagaop2024}, but not to the one in, e.g., \cite{Diaz-Fernandez-2017}, where the role of the negative eigenvalues (and their related eigenstates) is simply neglected. We show how to decompose (non uniquely) the Hilbert space where the model is defined in two orthogonal subspaces and how to construct coherent states in both these spaces. Also, in view of our special structure, it turns out that what is a {\em lowering operator} in one subspace (since it {\em moves toward the vacuum}), is a {\em raising operator} in the orthogonal space, since it {\em moves away from the vacuum}, and vice-versa. This aspect will be discussed in details and clarified all along Section \ref{sect2}. In Section \ref{subsect23} we restate essentially the same results in terms of tensor product Hilbert spaces, since this approach will simplify quite a bit what discussed later in Section \ref{sect3}.

Section \ref{sect3} is focused on the case of $V\neq0$, i.e. on the analysis of the effect of a chemical potential in our system. We discuss first the case of small $V$ ($V\in[0,1[$). We introduce  two different families of bicoherent states: those which are closer to the standard coherent states (meaning with this that they differ from a standard coherent state just because the orthonormal basis in terms of which the coherent state is expanded is replaced by two biorthonormal sets), and a second family in which, other than the previous difference, we {\bf also} replace the standard coefficient $\sqrt{n!}$ in the expansion\footnote{This is the standard terms one meets when dealing with the expansion of an coherent states of, say, an harmonic oscillator.} with something different, related to the eigenvalues of the non self-adjoint Hamiltonian of graphene, in presence of the chemical potential. These latter are connected to ladder operators which are {\em particularly interesting}, since they can be used to factorize the Hamiltonian. The properties of both these families of states will be analyzed in some details. In Section \ref{sect4b} we discuss the case $V>1$, which, as we shall see, leads to a larger $\mathcal{P}\mathcal{T}$ broken symmetry region and deserves an appropriate approach for the construction of the coherent states.

The paper is closed by Section \ref{sect4}, which contains our conclusions and plans for the future.

\section{Coherent states for graphene with $V=0$}\label{sect2}

We begin our analysis considering the case in which there is no chemical potential. To make the paper self-contained, it is useful to begin by reviewing some known results on the orthonormal basis (ONB) for our system. After this short introduction we will propose our definition of coherent states.

Graphene is a single layer of carbon atoms arranged in a hexagonal honeycomb lattice, which is the basic structural element of other allotropes including graphite, charcoal, carbon nanotubes
and fullerenes. 
Graphene is a zero--gap semiconductor, because its conduction and valence bands meet at
the Dirac points which are six locations in momentum space, on the edge of the Brillouin
zone, divided into two non-equivalent sets of three points, typically labeled as $K$ and $K'$.

We consider a layer of graphene in an external constant magnetic field along $z:{ \bf B} = -B \widehat{e}_3,$
which can be deduced from ${ \bf B} = \nabla \wedge {\bf A}$ with a vector potential in the symmetric gauge,
${\bf A} = (B/2)(y, -x, 0)$. The Hamiltonian for the two Dirac points $K$ and $K'$, that is the wavenumbers where  the energy eigenvalues can degenerate to zero, can be written as in \cite{Novoselov-2007}:
\be \label{01}
H_D=  \begin{pmatrix}
	H_K & 0 \\
	0 & H_{K'}
\end{pmatrix},
	\en
where, in units $\hslash = c = 1$, we have:
\be \label{02}
H_K = v_F\, {\bf{\sigma}\cdot ( \bf{p}} +e{\bf A} )=v_F  \begin{pmatrix}
	0 & \hat p_x - i \hat p_y + \frac{e B}{2}(\hat y + i \hat x) \\
	\hat p_x + i \hat p_y + \frac{e B}{2}(\hat y - i \hat x) & 0
\end{pmatrix}.
	\en
Here $\bf{p}=-i\,\bf{\nabla}=-i\,(\partial_x,\partial_y)$, $\sigma=(\sigma_x,\sigma_y)$ are the Pauli's matrices, and $H_{K'}$ is just the transpose of $H_K$. In (\ref{02}) $\hat x, \hat y, \hat p_x$ and $ \hat p_y$ are the canonical, Hermitian, two--dimensional position and momentum operators, which satisfy $[\hat x, \hat p_x ] = [\hat y, \hat p_y ] = i \1$, with all the
other commutators being zero, and where $\1$ is the identity operator in the Hilbert space $\Hil= \Lc^2(\mathbb{R}^2)$. The factor $v_F$ is the so--called Fermi velocity.

It is convenient to rescale these operators as follows

\be \label{04}
\hat X=\frac{1}{\xi}\hat x, \qquad \hat Y=\frac{1}{\xi}\hat{y}, \qquad \hat P_X= \xi \hat p_x \qquad \text{and} \qquad \hat P_Y= \xi \hat p_y,
\en
in which $\xi=\sqrt{2/(e|B|)}$ is the so--called magnetic length, and then to introduce
$ a_X = \dfrac{\hat X +i\hat P_X}{\sqrt{2}}$ and $ a_Y = \dfrac{\hat Y +i\hat P_Y}{\sqrt{2}} $. These are bosonic operators, as well as their combinations
\be \label{03}
{A_1} = \dfrac{ a_X -i  a_Y}{\sqrt{2}} \qquad \text{and} \qquad {A_2} =\dfrac{ a_X + i  a_Y}{\sqrt{2}}.
\en
Indeed, the following canonical commutation rules (CCRs) are satisfied:

\be \label{05}
[ a_X,  a_X^\dagger]=[ a_Y, a_Y^\dagger]=[{A_1},{A_1}^\dagger]=[{A_2},{A_2}^\dagger]=\1,
\en
and the Hamiltonian $H_K$ can be rewritten as:

\be \label{06}
H_K^{+} = \dfrac{2 i v_F}{\xi}   \begin{pmatrix}
	0 & {A_2}^\dagger \\
	-{A_2} & 0
\end{pmatrix}  \qquad \text{for} \quad B>0,
\en 
and

\be \label{07}
 H_K^{-} = \dfrac{2 i v_F}{\xi}    \begin{pmatrix}
	0 & -{A_1} \\
	{A_1}^\dagger & 0
\end{pmatrix} \qquad \text{for} \quad B<0.
\en
Of course, $H_{K'}^{+}$ depends only on ${A_2}$ and $H_{K'}^{-}$ on ${A_1}$.
In order to deal with an Hamiltonian of the kind \eqref{06} or \eqref{07} it is convenient to work in a new Hilbert space, $\Hil_2 = \Hil \oplus \Hil$, with scalar product $\langle f,g \rangle_2:= \langle f_1,g_1 \rangle + \langle f_2,g_2 \rangle$, where $f=\begin{pmatrix}
	f_1 \\
	f_2
\end{pmatrix} $ and $g=\begin{pmatrix}
g_1 \\
g_2
\end{pmatrix} $, 
with
 $f_1, f_2, g_1$ and $g_2$ in $\Hil$, and where $\langle.,.\rangle$ is the scalar product in $\Hil$.
We can introduce the vectors 

\be \label{08}
v_{n_1,0}= v_{n_1,0}^{+}=v_{n_1,0}^{-}=\begin{pmatrix}
 e_{n_1,0} \\
0
\end{pmatrix} 
\en
while, for $n_2 \geq 1$,
\be \label{09} 
v_{n_1,n_2}^{\pm}= \dfrac{1}{\sqrt{2}} \begin{pmatrix}
 e_{n_1,n_2} \\
\mp i e_{n_1,n_2-1}
\end{pmatrix} 
\en 
Here  $e_{n_1,n_2}:= \dfrac{1}{\sqrt{n_1! n_2!} } ({A_1}^\dagger)^{n_1} ({A_2}^\dagger)^{n_2} e_{0,0} $, where $e_{0,0}$ is the non zero {\it vacuum} of ${A_1}$ and ${A_2}$,
$$
{A_1}e_{0,0}={A_2}e_{0,0}=0.
$$
Incidentally we observe that this function can be explicitly computed, using the fact that $e_{0,0}$ must also satisfy $ a_X\,e_{0,0}= a_Y\, e_{0,0}=0$, so that
$$
e_{0,0}\longrightarrow e_{0,0}(X,Y)=\frac{1}{\sqrt{\pi}}e^{-\frac{1}{2}(X^2+Y^2)}.
$$
The set $\mathcal{V}_2=\{v_{n_1,n_2}^{
	k}, n_1\geq0, n_2 \geq1, k= \pm  \}\cup\{v_{n_1,0}, n_1\geq0\}$ is an ONB for $\Hil_2$, and its vectors are the eigenvectors of $H_K^{+}$:

\be \label{eigv}
H_K^{+}  v_{n_1,n_2}^{\pm} = E_{n_1,n_2}^{\pm} v_{n_1,n_2}^{\pm},
\en
with eigenvalues $ E_{n_1,n_2}^{\pm}= \pm \dfrac{2 v_f}{\xi} \sqrt{n_2}$, and
\be
\langle v_{n_1,n_2}^{
	k}, v_{m_1,m_2}^{
	j} \rangle_2=\delta_{n_1,m_1}\,\delta_{n_2,m_2}\,\delta_{k,j},
\label{orto}\en
$n_i,m_i,\geq0$, $i=1,2$, and $k,j=\pm$. If in the left-hand side of (\ref{orto}) $n_2=m_2=0$, then we identify $v_{n_1,0}^k$ and $v_{n_1,0}^j$ as in (\ref{08}). 
One can adapt these results to the other Hamiltonians $H_{K}^{-}, H_{K'}^{+}$ or $H_{K'}^{-}$.

\vspace{2mm}

{\bf Remark:--} In \cite{BagaHatano_2016} the set $\mathcal{V}_2$ was slightly different from the one considered here. Indeed it was defined as $\mathcal{V}_2=\{v_{n_1,n_2}^{
	k}, n_1\geq0, n_2 \geq0, k= \pm  \}$ which differs from ours because the vectors $v_{n_1,0}$ is counted twice. However, as a set, the two sets are clearly indistinguishable.

\vspace{2mm}

We see from (\ref{eigv}), and from the expression of $ E_{n_1,n_2}^{\pm}$, that $H_K^+$ (to which we will restrict, from now on) is unbounded from above {\bf and} from below, and that each energetic level has an infinite degeneracy, since $ E_{n_1,n_2}^{\pm}$ does not depend on $n_1$. In what follows it is convenient to use a slightly different notation, by introducing the following vectors:
\be
c_{n,p}=\left\{
\begin{array}{ll}
	v_{n,p}^{
		+}, \hspace{2cm} p\geq1,\\
	v_{n,0}, \hspace{2cm} p=0,\\
		v_{n,-p}^-,\hspace{1.8cm} p\leq-1,
\end{array}
\right.
\label{f1}\en
where $n\geq0$.
Hence the set $\V_2$ can be rewritten as $\V_2=\{c_{n,p},\, n\geq0, p\in\mathbb{Z}\}$, and
\be
H_K^+c_{n,p}=\E_{n,p}c_{n,p}, \qquad \E_{n,p}=\text{sign}(p)\,\frac{2v_F}{\xi}\sqrt{|p|},
\label{f2}\en
where $n\geq0$ while $p\in\mathbb{Z}$. Of course we have
\be
\langle c_{n,p}, c_{m,q}\rangle_2=\delta_{n,m}\delta_{p,q},
\label{f3}\en
$\forall n,m\geq0$ and $\forall p,q\in\mathbb{Z}$.
If we now introduce the following $p$-depending Hilbert spaces
$$
\Hil_2(p)=\overline{l.s.\{c_{n,p},\,n\geq0\}}^{\|.\|},
$$
$p\in\mathbb{Z}$, it follows that $\Hil_2=\oplus_{p\in\mathbb{Z}}\Hil_2(p)$. Each $\Hil_2(p)$ corresponds to a different energetic level of $H_K^+$, and all the levels are mutually orthogonal: if $f\in\Hil_2(p)$ and $g\in\Hil_2(q)$, $p,q\in\mathbb{Z}$ with $p\neq q$, then $\langle f,g\rangle_2=0$.

\subsection{Coherent states}\label{subsect21}

To construct ordinary coherent states, see \cite{gazeaubook} for instance, one usually construct series of vectors starting from a certain {\em vacuum}, i.e. a vector which is annihilated by a lowering operator. For instance, if we have two bosonic operators $c$ and $c^\dagger$, we construct such a family of vectors looking first for a vector $e_0\neq0$ satisfying $ce_0=0$, and then we define $e_n=\frac{{c^\dagger}^n}{\sqrt{n!}}\,e_0$, $n\geq0$. Hence a coherent state is the normalized vector
$$
\Phi(z)=e^{-|z|^2/2}\sum_{n=0}^\infty\,\frac{z^n}{\sqrt{n!}}\,e_n,
$$
where $z\in\mathbb{C}$ and the number $n\geq0$ in $\sqrt{n!}$ is directly related to the eigenvalue of the Hamiltonian of the shifted harmonic oscillator $H=\omega c^\dagger c$, with $\omega=1$. We refer to \cite{gazeaubook,didier,bagspringer} and references therein, for several properties of $\Phi(z)$. What is relevant for us is that, in our case, we cannot use the same approach since the set of eigenvalues of $H_K^+$ is not bounded from below, see (\ref{f2}). In the attempt of defining coherent states also in this situation one could first simply think to replace the sum $\sum_{n=0}^\infty$ with $\sum_{n=-\infty}^\infty$ in a possible new version of $\Phi(z)$. Of course, this creates some difference when exploring the existence of this new vector, since negative powers of $z$ also appear in the expansion. Hence this new state cannot be defined in all of $\mathbb{C}$, since we have to find the convergent region of a Laurent series. Indeed, if we consider $\Psi(z)=N(z)\sum_{n=-\infty}^\infty\,\frac{z^n}{\sqrt{|n|!}}\,e_n$, $N(z)$ being  a suitable normalization, we would have $\|\Psi(z)\|^2=|N(z)|^2\sum_{n=-\infty}^\infty\,\frac{|z|^{2n}}{|n|!}$, which is a Laurent (and not a Taylor) series. But this is not really a major problem, in our opinion: we know that {\em nonlinear coherent states} don't need to be defined in all of $\mathbb{C}$, see \cite{gazeaubook} for instance. The most serious problem is rather the fact that $\Psi(z)$ cannot be the eigenstate of $c$ with eigenvalue $z$, as it is easily seen. So a different definition of our coherent state should be proposed. This is exactly what was done recently in \cite{bagaop2024}, where the idea was  {\em to double} the original Hilbert space where $\Phi(z)$ is defined, and to work with {\em vector coherent states}. In this way, however, the final result is that we have to work with a special subset of $\Hil_2$. Here we follow a different strategy, working in the original space $\Hil_2$ as much as possible. Incidentally we observe that, as already pointed out in (\ref{f1}), we need to deal with two indexes, $n$ and $p$, having different ranges of values.

We start  introducing the operator $\Ac_1$ on $\Hil_2$ as follows:
\be
\Ac_1=\begin{pmatrix}
	A_1 & 0 \\
	0 & A_1
\end{pmatrix},
\label{cs1}\en
which together with its adjoint $\Ac_1^\dagger$ satisfies the CCR $[\Ac_1,\Ac_1^\dagger]=\1_2$, where $\1_2$ is the identity operator on $\Hil_2$. It is easy to check that
\be\Ac_1c_{n,p}=\sqrt{n}\,c_{n-1,p}, \qquad \mbox{ with  }\qquad  \Ac_1c_{0,p}=0,
\label{cs2}\en
for all $p\in\mathbb{Z}$ and $n\geq1$. In view of the different nature of $n$ and $p$ in $c_{n,p}$ it is not convenient to introduce ladder operators acting on the second index, $p$ as in (\ref{cs1}). We rather observe that $\Ac_1$ can be rewritten as
$$
\Ac_1=\sum_{n=0}^\infty\sum_{p=-\infty}^\infty\sqrt{n+1}\,|c_{n,p}\rangle_2 {_2\langle} c_{n+1,p}|,
$$
which is densely defined on $\Hil_2$ since its domain, $D(\Ac_1)$, contains the linear span of all the $c_{n,p}$, $\Lc_c$, which is dense in $\Hil_2$ since $\Hil_2=\overline{\Lc_c}^{\|.\|}$.

With this in mind, we introduce a new operator $\Ac_2$ as follows:
\be
\Ac_2=\sum_{n=0}^\infty\sum_{p=-\infty}^\infty\sqrt{|p+1|}\,|c_{n,p}\rangle_2 {_2\langle} c_{n,p+1}|.
\label{cs3}\en
As before, $D(\Ac_2)\supseteq\Lc_c$. Hence $\Ac_2$ is also densely defined. Moreover, it satisfies the following lowering equation:
\be\Ac_2c_{n,p}=\sqrt{|p|}\,c_{n,p-1}, 
\label{cs4}\en
for all $n\geq0$ and $p\in\mathbb{Z}$. In particular, this means that $\Ac_2c_{n,0}=0$, $\forall n\geq0$, but $\Ac_2$ can also act  on vectors $c_{n,p}$ with $p\leq-1$, returning a non zero result. This means that $c_{n,0}$ is not (are not, actually, due to the presence of $n$) a {\em ground state}: each $c_{n,0}$ is simply a vector which is annihilated by $\Ac_2$. It is not hard to find the adjoint of $\Ac_2$, which satisfies the following raising equation;
\be
\Ac_2^\dagger c_{n,p}=\sqrt{|p+1|}\,c_{n,p+1}, 
\label{cs5}\en
for all $n\geq0$ and $p\in\mathbb{Z}$. We can write
\be
\Ac_2^\dagger=\sum_{n=0}^\infty\sum_{p=-\infty}^\infty\sqrt{|p+1|}\,|c_{n,p+1}\rangle_2 {_2\langle} c_{n,p}|,
\label{cs6}\en
and $D(\Ac_2^\dagger)\supseteq\Lc_c$. It is interesting to notice that $\Ac_2^\dagger$ acts as a raising operator on the various $c_{n,p}$, except for $c_{n,-1}$, which is annihilated by $\Ac_2^\dagger$: $\Ac_2^\dagger c_{n,-1}=0$, $\forall n\geq0$.

\vspace{2mm}

{\bf Remark:--} We observe that
\be
[H_K^+,\Ac_2^\dagger\Ac_2]f=0,
\label{cs7}\en
$\forall f\in\Lc_c$. This follows from (\ref{f2}), (\ref{cs2}) and (\ref{cs5}). However, while the eigenvalues of $H_K^+$ depends on $p$ as in $\sqrt{|p|}$, those of $\Ac_2^\dagger\Ac_2$ depends on $p$ as $|p|$. Moreover, while $\Ac_2^\dagger\Ac_2$ is a positive operator, $H_K^+$ is not. Hence it is clear that $H_K^+\neq\Ac_2^\dagger\Ac_2$: $H_k^+$ is not factorizable in terms of $\Ac_2$ and $\Ac_2^\dagger$.

\vspace{2mm}

Let us now introduce the following subspaces of $\Hil_2$: $\Hil_2^+=\oplus_{n=0}^\infty\Hil_2(n)$, and $\Hil_2^-=\oplus_{n=-\infty}^{-1}\Hil_2(n)$. It is clear that $\Hil_2=\Hil_2^+\oplus\Hil_2^-$ and that, taken $f\in\Hil_2^+$ and $g\in\Hil_2^-$, $\langle f, g\rangle_2=0$.  These two spaces are interesting in view of their behavior with respect $\Ac_2$ and $\Ac_2^\dagger$: they are {\em disconnected}: if we act on some $f\in\Hil_2^+$ with $\Ac_2$ or with $\Ac_2^\dagger$, and with their powers, we get some other vector which is still in $\Hil_2^+$. The same happens if we start from a vector $g\in\Hil_2^-$.  The situation is described in Figure \ref{fig2}a.


Because of formulas (\ref{cs4}) and (\ref{cs5}), and of what we have discussed above, we could interpret $c_{n,0}$ as a {\em quasi-vacuum} for $\Ac_2$, and $c_{n,-1}$ as a {\em quasi-vacuum} for $\Ac_2^\dagger$. This suggests us to define the following vectors:
\be
\Phi_\Ac^+(z_1,z_2)=e^{-(|z_1|^2+|z_2|^2)/2}\sum_{n1=0}^\infty\sum_{n2=0}^\infty\frac{z_1^{n_1}\,z_2^{n_2}}{\sqrt{n_1!\,n_2!}}\,c_{n_1,n_2},
\label{cs8}\en
and
\be
\Phi_\Ac^-(z_1,z_2)=e^{-(|z_1|^2+|z_2|^2)/2}\sum_{n1=0}^\infty\sum_{n2=0}^\infty\frac{z_1^{n_1}\,z_2^{n_2}}{\sqrt{n_1!\,n_2!}}\,c_{n_1,-n_2-1},
\label{cs9}\en
which live respectively in $\Hil_2^+$ and $\Hil_2^-$ and, as such, are mutually orthogonal: $$\langle \Phi_\Ac^+(z_1,z_2), \Phi_\Ac^-(z_1,z_2)\rangle_2=0,$$ for all $z_1, z_2\in\mathbb{C}$. Moreover, it is also possible to check (and it is not a surprise, of course!) that the two series in (\ref{cs8}) and (\ref{cs9}) converge in the entire complex plane, and that
\be
\|\Phi_\Ac^+(z_1,z_2)\|_2=\|\Phi_\Ac^-(z_1,z_2)\|_2=1,
\label{cs10}\en
$\forall z_1, z_2\in\mathbb{C}$. For all these $z_1$ and $z_2$ it is further possible to prove the following eigenvalue equations:
\be
\Ac_1\Phi_\Ac^+(z_1,z_2)=z_1\,\Phi_\Ac^+(z_1,z_2), \qquad \Ac_1\Phi_\Ac^-(z_1,z_2)=z_1\,\Phi_\Ac^-(z_1,z_2),
\label{cs11}\en
as well as
\be
\Ac_2\Phi_\Ac^+(z_1,z_2)=z_2\,\Phi_\Ac^+(z_1,z_2), \qquad \Ac_2^\dagger\Phi_\Ac^-(z_1,z_2)=z_2\,\Phi_\Ac^-(z_1,z_2).
\label{cs12}\en
In particular, the equations in (\ref{cs12}) are in agreement with our interpretation of $\Ac_2$ as a lowering operator in $\Hil_2^+$, and of $\Ac_2^\dagger$ as a different lowering operator (meaning with this that $\Ac_2^\dagger$ {\em moves the state} toward its own quasi vacuum, $c_{n,-1}$) on $\Hil_2^-$. The evident difference between (\ref{cs11}) and (\ref{cs12})  is in the presence of only $\Ac_1$ in the first equation, and of both $\Ac_2$ and $\Ac_2^\dagger$ in the second. This is again due to the different ranges of $n$ and $p$, and on their role, in our construction.

To conclude our analysis of these states, we can also check that they satisfy the following resolutions of the identity:
\be
\langle f_+, g_+\rangle_2=\int_{\mathbb{C}}\frac{d^2z_1}{\pi}\int_{\mathbb{C}}\frac{d^2z_2}{\pi}\langle f_+, \Phi_\Ac^+(z_1,z_2)\rangle_2\langle \Phi_\Ac^+(z_1,z_2), g_+\rangle_2,
\label{cs13}\en
for all $f_+,g_+\in\Hil_2^+$, and 
\be
\langle f_-, g_-\rangle_2=\int_{\mathbb{C}}\frac{d^2z_1}{\pi}\int_{\mathbb{C}}\frac{d^2z_2}{\pi}\langle f_-, \Phi_\Ac^-(z_1,z_2)\rangle_2\langle \Phi_\Ac^-(z_1,z_2), g_-\rangle_2,
\label{cs14}\en
for all $f_-,g_-\in\Hil_2^-$.

 One might wonder if the normalized vector $\Phi_\Ac(z_1,z_2)=\frac{1}{\sqrt{2}}(\Phi_\Ac^+(z_1,z_2)+\Phi_\Ac^-(z_1,z_2))$ produces a resolution of the identity in all of $\Hil_2$. The answer is negative. Indeed, if we take two vectors $f,g\in\Hil_2$, we can write $f=f_++f_-$ and $g=g_++g_-$, where $f_+,g_+\in\Hil_2^+$ and $f_-,g_-\in\Hil_2^-$. Then we have, with easy computations,
$$
\int_{\mathbb{C}}\frac{d^2z_1}{\pi}\int_{\mathbb{C}}\frac{d^2z_2}{\pi}\langle f, \Phi_\Ac(z_1,z_2)\rangle_2\langle \Phi_\Ac(z_1,z_2), g\rangle_2=\frac{1}{2}\langle f, g\rangle_2+
$$
$$
+\frac{1}{2}\int_{\mathbb{C}}\frac{d^2z_1}{\pi}\int_{\mathbb{C}}\frac{d^2z_2}{\pi}\left(\langle f_+, \Phi_\Ac^+(z_1,z_2)\rangle_2\langle \Phi_\Ac^-(z_1,z_2),g_-\rangle_2+\langle f_-, \Phi_\Ac^-(z_1,z_2)\rangle_2\langle \Phi_\Ac^+(z_1,z_2),g_+\rangle_2\right),
$$
which is in general different from $\langle f, g\rangle_2$. Moreover, $\Phi_\Ac(z_1,z_2)$ is not an eigenstate of $\Ac_2$ or $\Ac_2^\dagger$.  Hence we conclude that ours are coherent states not in $\Hil_2$, but in two orthogonal subspaces of $\Hil_2$, where they possess all the standard properties of coherent states.

\subsection{A different decomposition of $\Hil_2$}\label{sect22}

One might wonder why we should be satisfied with the decomposition $\Hil_2=\Hil_2^+\oplus\Hil_2^-$ we have considered above. Indeed this is not the only possibility, but it is the natural one if we work with the operators $\Ac_2$ and $\Ac_2^\dagger$. In fact, we could easily modify what discussed so far by replacing these operators with, e.g., two different ones, $\Bc_2$ and $\Bc_2^\dagger$, defined as follows:

\be
\Bc_2=\sum_{n=0}^\infty\sum_{p=-\infty}^\infty\sqrt{|p|}\,|c_{n,p+1}\rangle_2 {_2\langle} c_{n,p}|,
\label{cs15}\en
so that
\be\Bc_2c_{n,p}=\sqrt{|p|}\,c_{n,p+1}, 
\label{cs16}\en
and
\be
\Bc_2^\dagger=\sum_{n=0}^\infty\sum_{p=-\infty}^\infty\sqrt{|p|}\,|c_{n,p}\rangle_2 {_2\langle} c_{n,p+1}|,
\label{cs17}\en
so that
\be\Bc_2^\dagger c_{n,p}=\sqrt{|p-1|}\,c_{n,p-1}, 
\label{cs18}\en
$\forall n\geq0$ and $p\in\mathbb{Z}$. In particular, these equations imply that both $\Bc_2$ and $\Bc_2^\dagger$ are densely defined (at least on $\Lc_c$), and that
\be
\Bc_2\,c_{n,0}=0, \qquad \Bc_2^\dagger c_{n,1}=0, \qquad n\geq0.
\label{cs19}\en
We find, similarly to what has been observed before, that $[H_K^+,\Bc_2\Bc_2^\dagger]f=0$, $\forall f\in\Lc_c$. Notice that, here, $\Bc_2$ behaves as a raising operator on $\Hil_2$, while  $\Bc_2^\dagger$ acts as a lowering operator on $\Hil_2$. For these operators, the {\em quasi-vacua} are  $c_{n,0}$ and $c_{n,1}$, see (\ref{cs19}). The situation is shown in Figure \ref{fig2}, where the two different decompositions of $\Hil_2$ arising from our approach are shown:

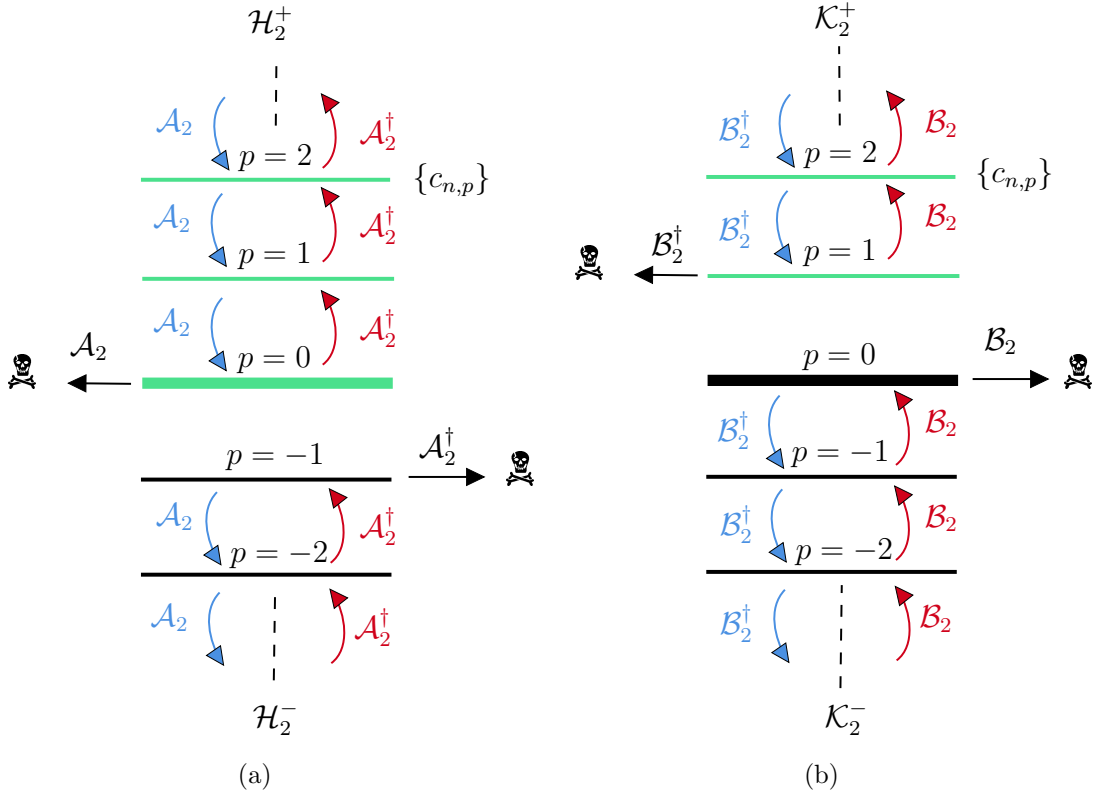
\begin{figure}[!ht]
	\begin{center}
	\subfigure[]{

	\tikzset{every picture/.style={line width=0.75pt}} 
	
	\begin{tikzpicture}[x=0.75pt,y=0.75pt,yscale=-1,xscale=1]
		
		\draw [color={rgb, 255:red, 74; green, 224; blue, 140 }  ,draw opacity=1 ][line width=1.5]    (237,191) -- (364,191) ;
		\draw [color={rgb, 255:red, 74; green, 224; blue, 140 }  ,draw opacity=1 ][line width=1.5]    (236.33,141) -- (362,141) ;
		\draw [color={rgb, 255:red, 208; green, 2; blue, 27 }  ,draw opacity=1 ]   (328,182.33) .. controls (337.81,174.97) and (335.21,157.46) .. (329.02,147.43) ;
		\draw [shift={(327.33,145)}, rotate = 51.84] [fill={rgb, 255:red, 208; green, 2; blue, 27 }  ,fill opacity=1 ][line width=0.08]  [draw opacity=0] (8.93,-4.29) -- (0,0) -- (8.93,4.29) -- cycle    ;
		\draw [color={rgb, 255:red, 208; green, 2; blue, 27 }  ,draw opacity=1 ]   (328,135) .. controls (337.81,127.64) and (335.21,110.12) .. (329.02,100.1) ;
		\draw [shift={(327.33,97.67)}, rotate = 51.84] [fill={rgb, 255:red, 208; green, 2; blue, 27 }  ,fill opacity=1 ][line width=0.08]  [draw opacity=0] (8.93,-4.29) -- (0,0) -- (8.93,4.29) -- cycle    ;
		\draw [color={rgb, 255:red, 74; green, 144; blue, 226 }  ,draw opacity=1 ]   (279.06,133.45) .. controls (273.72,124.21) and (269.57,108.74) .. (278.67,99.33) ;
		\draw [shift={(280.67,136)}, rotate = 235.3] [fill={rgb, 255:red, 74; green, 144; blue, 226 }  ,fill opacity=1 ][line width=0.08]  [draw opacity=0] (8.93,-4.29) -- (0,0) -- (8.93,4.29) -- cycle    ;
		\draw [color={rgb, 255:red, 74; green, 144; blue, 226 }  ,draw opacity=1 ]   (277.72,182.12) .. controls (272.38,172.88) and (268.23,157.4) .. (277.33,148) ;
		\draw [shift={(279.33,184.67)}, rotate = 235.3] [fill={rgb, 255:red, 74; green, 144; blue, 226 }  ,fill opacity=1 ][line width=0.08]  [draw opacity=0] (8.93,-4.29) -- (0,0) -- (8.93,4.29) -- cycle    ;
		\draw [color={rgb, 255:red, 74; green, 224; blue, 140 }  ,draw opacity=1 ][line width=4.25]    (237,243.67) -- (363.33,243.67) ;
		\draw [color={rgb, 255:red, 208; green, 2; blue, 27 }  ,draw opacity=1 ]   (328,234.33) .. controls (337.81,226.97) and (335.21,209.46) .. (329.02,199.43) ;
		\draw [shift={(327.33,197)}, rotate = 51.84] [fill={rgb, 255:red, 208; green, 2; blue, 27 }  ,fill opacity=1 ][line width=0.08]  [draw opacity=0] (8.93,-4.29) -- (0,0) -- (8.93,4.29) -- cycle    ;
		\draw [color={rgb, 255:red, 74; green, 144; blue, 226 }  ,draw opacity=1 ]   (277.72,234.79) .. controls (272.38,225.55) and (268.23,210.07) .. (277.33,200.67) ;
		\draw [shift={(279.33,237.33)}, rotate = 235.3] [fill={rgb, 255:red, 74; green, 144; blue, 226 }  ,fill opacity=1 ][line width=0.08]  [draw opacity=0] (8.93,-4.29) -- (0,0) -- (8.93,4.29) -- cycle    ;
		\draw [color={rgb, 255:red, 0; green, 0; blue, 0 }  ,draw opacity=1 ][line width=1.5]    (236.33,340.33) -- (362.67,340.33) ;
		\draw [color={rgb, 255:red, 0; green, 0; blue, 0 }  ,draw opacity=1 ][line width=1.5]    (236.33,292.33) -- (362.67,292.33) ;
		\draw [color={rgb, 255:red, 208; green, 2; blue, 27 }  ,draw opacity=1 ]   (332,334.33) .. controls (341.81,326.97) and (339.21,309.46) .. (333.02,299.43) ;
		\draw [shift={(331.33,297)}, rotate = 51.84] [fill={rgb, 255:red, 208; green, 2; blue, 27 }  ,fill opacity=1 ][line width=0.08]  [draw opacity=0] (8.93,-4.29) -- (0,0) -- (8.93,4.29) -- cycle    ;
		\draw  [dash pattern={on 4.5pt off 4.5pt}]  (304.33,113.67) -- (304.67,79) ;
		\draw [color={rgb, 255:red, 74; green, 144; blue, 226 }  ,draw opacity=1 ]   (275.06,332.79) .. controls (269.72,323.55) and (265.57,308.07) .. (274.67,298.67) ;
		\draw [shift={(276.67,335.33)}, rotate = 235.3] [fill={rgb, 255:red, 74; green, 144; blue, 226 }  ,fill opacity=1 ][line width=0.08]  [draw opacity=0] (8.93,-4.29) -- (0,0) -- (8.93,4.29) -- cycle    ;
		\draw  [dash pattern={on 4.5pt off 4.5pt}]  (303.33,393.67) -- (304,347) ;
		\draw    (231.33,244) -- (202.67,243.7) ;
		\draw [shift={(199.67,243.67)}, rotate = 0.6] [fill={rgb, 255:red, 0; green, 0; blue, 0 }  ][line width=0.08]  [draw opacity=0] (8.93,-4.29) -- (0,0) -- (8.93,4.29) -- cycle    ;
		\draw [color={rgb, 255:red, 208; green, 2; blue, 27 }  ,draw opacity=1 ]   (332.33,385) .. controls (342.15,377.64) and (339.55,360.12) .. (333.36,350.1) ;
		\draw [shift={(331.67,347.67)}, rotate = 51.84] [fill={rgb, 255:red, 208; green, 2; blue, 27 }  ,fill opacity=1 ][line width=0.08]  [draw opacity=0] (8.93,-4.29) -- (0,0) -- (8.93,4.29) -- cycle    ;
		\draw [color={rgb, 255:red, 74; green, 144; blue, 226 }  ,draw opacity=1 ]   (276.39,383.45) .. controls (271.05,374.21) and (266.9,358.74) .. (276,349.33) ;
		\draw [shift={(278,386)}, rotate = 235.3] [fill={rgb, 255:red, 74; green, 144; blue, 226 }  ,fill opacity=1 ][line width=0.08]  [draw opacity=0] (8.93,-4.29) -- (0,0) -- (8.93,4.29) -- cycle    ;
		\draw    (371.33,291) -- (406.5,291) ;
		\draw [shift={(409.5,291)}, rotate = 180] [fill={rgb, 255:red, 0; green, 0; blue, 0 }  ][line width=0.08]  [draw opacity=0] (8.93,-4.29) -- (0,0) -- (8.93,4.29) -- cycle    ;
		
		\draw (284,171.07) node [anchor=north west][inner sep=0.75pt]    {$p=1$};
		\draw (284.33,121.73) node [anchor=north west][inner sep=0.75pt]    {$p=2$};
		\draw (345,153.07) node [anchor=north west][inner sep=0.75pt]  [color={rgb, 255:red, 208; green, 2; blue, 27 }  ,opacity=1 ]  {$\mathcal{A}_{2}^{\dagger }$};
		\draw (345,105.73) node [anchor=north west][inner sep=0.75pt]  [color={rgb, 255:red, 208; green, 2; blue, 27 }  ,opacity=1 ]  {$\mathcal{A}_{2}^{\dagger }$};
		\draw (242.33,103.73) node [anchor=north west][inner sep=0.75pt]  [color={rgb, 255:red, 74; green, 144; blue, 226 }  ,opacity=1 ]  {$\mathcal{A}_{2}$};
		\draw (241,152.4) node [anchor=north west][inner sep=0.75pt]  [color={rgb, 255:red, 74; green, 144; blue, 226 }  ,opacity=1 ]  {$\mathcal{A}_{2}$};
		\draw (283.67,223.73) node [anchor=north west][inner sep=0.75pt]    {$p=0$};
		\draw (345,205.07) node [anchor=north west][inner sep=0.75pt]  [color={rgb, 255:red, 208; green, 2; blue, 27 }  ,opacity=1 ]  {$\mathcal{A}_{2}^{\dagger }$};
		\draw (241,205.07) node [anchor=north west][inner sep=0.75pt]  [color={rgb, 255:red, 74; green, 144; blue, 226 }  ,opacity=1 ]  {$\mathcal{A}_{2}$};
		\draw (280,322.07) node [anchor=north west][inner sep=0.75pt]    {$p=-2$};
		\draw (278,272.73) node [anchor=north west][inner sep=0.75pt]    {$p=-1$};
		\draw (345,305.07) node [anchor=north west][inner sep=0.75pt]  [color={rgb, 255:red, 208; green, 2; blue, 27 }  ,opacity=1 ]  {$\mathcal{A}_{2}^{\dagger }$};
		\draw (375,264.4) node [anchor=north west][inner sep=0.75pt]    {$\mathcal{A}_{2}^{\dagger }$};
		\draw (242.33,303.07) node [anchor=north west][inner sep=0.75pt]  [color={rgb, 255:red, 74; green, 144; blue, 226 }  ,opacity=1 ]  {$\mathcal{A}_{2}$};
		\draw (199,217.07) node [anchor=north west][inner sep=0.75pt]    {$\mathcal{A}_{2}$};
		\draw (167,226.07) node [anchor=north west][inner sep=0.75pt]  [font=\normalsize]  {$\skull $};
		\draw (342.33,355.73) node [anchor=north west][inner sep=0.75pt]  [color={rgb, 255:red, 208; green, 2; blue, 27 }  ,opacity=1 ]  {$\mathcal{A}_{2}^{\dagger }$};
		\draw (239.67,353.73) node [anchor=north west][inner sep=0.75pt]  [color={rgb, 255:red, 74; green, 144; blue, 226 }  ,opacity=1 ]  {$\mathcal{A}_{2}$};
		\draw (418,275.07) node [anchor=north west][inner sep=0.75pt]  [font=\normalsize]  {$\skull $};
		\draw (288.67,52.07) node [anchor=north west][inner sep=0.75pt]    {$\mathcal{H}_{2}^{+}$};
		\draw (291.33,402.73) node [anchor=north west][inner sep=0.75pt]    {$\mathcal{H}_{2}^{-}$};
		\draw (372,131.4) node [anchor=north west][inner sep=0.75pt]    {$\{c_{n,p}\}$};

\end{tikzpicture}}
\subfigure[]{

\tikzset{every picture/.style={line width=0.75pt}} 

\begin{tikzpicture}[x=0.75pt,y=0.75pt,yscale=-1,xscale=1]
	
	\draw [color={rgb, 255:red, 74; green, 224; blue, 140 }  ,draw opacity=1 ][line width=1.5]    (238.33,152.33) -- (365.33,152.33) ;
	\draw [color={rgb, 255:red, 74; green, 224; blue, 140 }  ,draw opacity=1 ][line width=1.5]    (237.67,102.33) -- (363.33,102.33) ;
	\draw [color={rgb, 255:red, 208; green, 2; blue, 27 }  ,draw opacity=1 ]   (329.33,143.67) .. controls (339.15,136.31) and (336.55,118.79) .. (330.36,108.77) ;
	\draw [shift={(328.67,106.33)}, rotate = 51.84] [fill={rgb, 255:red, 208; green, 2; blue, 27 }  ,fill opacity=1 ][line width=0.08]  [draw opacity=0] (8.93,-4.29) -- (0,0) -- (8.93,4.29) -- cycle    ;
	\draw [color={rgb, 255:red, 208; green, 2; blue, 27 }  ,draw opacity=1 ]   (329.33,96.33) .. controls (339.15,88.97) and (336.55,71.46) .. (330.36,61.43) ;
	\draw [shift={(328.67,59)}, rotate = 51.84] [fill={rgb, 255:red, 208; green, 2; blue, 27 }  ,fill opacity=1 ][line width=0.08]  [draw opacity=0] (8.93,-4.29) -- (0,0) -- (8.93,4.29) -- cycle    ;
	\draw [color={rgb, 255:red, 74; green, 144; blue, 226 }  ,draw opacity=1 ]   (280.39,94.79) .. controls (275.05,85.55) and (270.9,70.07) .. (280,60.67) ;
	\draw [shift={(282,97.33)}, rotate = 235.3] [fill={rgb, 255:red, 74; green, 144; blue, 226 }  ,fill opacity=1 ][line width=0.08]  [draw opacity=0] (8.93,-4.29) -- (0,0) -- (8.93,4.29) -- cycle    ;
	\draw [color={rgb, 255:red, 74; green, 144; blue, 226 }  ,draw opacity=1 ]   (279.06,143.45) .. controls (273.72,134.21) and (269.57,118.74) .. (278.67,109.33) ;
	\draw [shift={(280.67,146)}, rotate = 235.3] [fill={rgb, 255:red, 74; green, 144; blue, 226 }  ,fill opacity=1 ][line width=0.08]  [draw opacity=0] (8.93,-4.29) -- (0,0) -- (8.93,4.29) -- cycle    ;
	\draw [color={rgb, 255:red, 0; green, 0; blue, 0 }  ,draw opacity=1 ][line width=4.25]    (238.33,205) -- (364.67,205) ;
	\draw [color={rgb, 255:red, 0; green, 0; blue, 0 }  ,draw opacity=1 ][line width=1.5]    (237.67,301.67) -- (364,301.67) ;
	\draw [color={rgb, 255:red, 0; green, 0; blue, 0 }  ,draw opacity=1 ][line width=1.5]    (237.67,253.67) -- (364,253.67) ;
	\draw [color={rgb, 255:red, 208; green, 2; blue, 27 }  ,draw opacity=1 ]   (333.33,295.67) .. controls (343.15,288.31) and (340.55,270.79) .. (334.36,260.77) ;
	\draw [shift={(332.67,258.33)}, rotate = 51.84] [fill={rgb, 255:red, 208; green, 2; blue, 27 }  ,fill opacity=1 ][line width=0.08]  [draw opacity=0] (8.93,-4.29) -- (0,0) -- (8.93,4.29) -- cycle    ;
	\draw  [dash pattern={on 4.5pt off 4.5pt}]  (305,76.33) -- (304.67,36.33) ;
	\draw [color={rgb, 255:red, 74; green, 144; blue, 226 }  ,draw opacity=1 ]   (276.39,294.12) .. controls (271.05,284.88) and (266.9,269.4) .. (276,260) ;
	\draw [shift={(278,296.67)}, rotate = 235.3] [fill={rgb, 255:red, 74; green, 144; blue, 226 }  ,fill opacity=1 ][line width=0.08]  [draw opacity=0] (8.93,-4.29) -- (0,0) -- (8.93,4.29) -- cycle    ;
	\draw  [dash pattern={on 4.5pt off 4.5pt}]  (305.33,360.33) -- (306,308.33) ;
	\draw    (233.33,152) -- (204.67,151.7) ;
	\draw [shift={(201.67,151.67)}, rotate = 0.6] [fill={rgb, 255:red, 0; green, 0; blue, 0 }  ][line width=0.08]  [draw opacity=0] (8.93,-4.29) -- (0,0) -- (8.93,4.29) -- cycle    ;
	\draw [color={rgb, 255:red, 208; green, 2; blue, 27 }  ,draw opacity=1 ]   (333.67,346.33) .. controls (343.48,338.97) and (340.88,321.46) .. (334.69,311.43) ;
	\draw [shift={(333,309)}, rotate = 51.84] [fill={rgb, 255:red, 208; green, 2; blue, 27 }  ,fill opacity=1 ][line width=0.08]  [draw opacity=0] (8.93,-4.29) -- (0,0) -- (8.93,4.29) -- cycle    ;
	\draw [color={rgb, 255:red, 74; green, 144; blue, 226 }  ,draw opacity=1 ]   (277.72,344.79) .. controls (272.38,335.55) and (268.23,320.07) .. (277.33,310.67) ;
	\draw [shift={(279.33,347.33)}, rotate = 235.3] [fill={rgb, 255:red, 74; green, 144; blue, 226 }  ,fill opacity=1 ][line width=0.08]  [draw opacity=0] (8.93,-4.29) -- (0,0) -- (8.93,4.29) -- cycle    ;
	\draw    (372,204.33) -- (407.17,204.33) ;
	\draw [shift={(410.17,204.33)}, rotate = 180] [fill={rgb, 255:red, 0; green, 0; blue, 0 }  ][line width=0.08]  [draw opacity=0] (8.93,-4.29) -- (0,0) -- (8.93,4.29) -- cycle    ;
	\draw [color={rgb, 255:red, 208; green, 2; blue, 27 }  ,draw opacity=1 ]   (333.33,247.67) .. controls (343.15,240.31) and (340.55,222.79) .. (334.36,212.77) ;
	\draw [shift={(332.67,210.33)}, rotate = 51.84] [fill={rgb, 255:red, 208; green, 2; blue, 27 }  ,fill opacity=1 ][line width=0.08]  [draw opacity=0] (8.93,-4.29) -- (0,0) -- (8.93,4.29) -- cycle    ;
	\draw [color={rgb, 255:red, 74; green, 144; blue, 226 }  ,draw opacity=1 ]   (275.06,246.79) .. controls (269.72,237.55) and (265.57,222.07) .. (274.67,212.67) ;
	\draw [shift={(276.67,249.33)}, rotate = 235.3] [fill={rgb, 255:red, 74; green, 144; blue, 226 }  ,fill opacity=1 ][line width=0.08]  [draw opacity=0] (8.93,-4.29) -- (0,0) -- (8.93,4.29) -- cycle    ;
	
	\draw (285.33,132.4) node [anchor=north west][inner sep=0.75pt]    {$p=1$};
	\draw (285.67,83.07) node [anchor=north west][inner sep=0.75pt]    {$p=2$};
	\draw (346.33,114.4) node [anchor=north west][inner sep=0.75pt]  [color={rgb, 255:red, 208; green, 2; blue, 27 }  ,opacity=1 ]  {$\mathcal{B}_{2}$};
	\draw (346.33,67.07) node [anchor=north west][inner sep=0.75pt]  [color={rgb, 255:red, 208; green, 2; blue, 27 }  ,opacity=1 ]  {$\mathcal{B}_{2}$};
	\draw (285,185.07) node [anchor=north west][inner sep=0.75pt]    {$p=0$};
	\draw (376.33,177.07) node [anchor=north west][inner sep=0.75pt]    {$\mathcal{B}_{2}$};
	\draw (281.33,283.4) node [anchor=north west][inner sep=0.75pt]    {$p=-2$};
	\draw (279.33,234.07) node [anchor=north west][inner sep=0.75pt]    {$p=-1$};
	\draw (346.33,266.4) node [anchor=north west][inner sep=0.75pt]  [color={rgb, 255:red, 208; green, 2; blue, 27 }  ,opacity=1 ]  {$\mathcal{B}_{2}$};
	\draw (169,134.07) node [anchor=north west][inner sep=0.75pt]  [font=\normalsize]  {$\skull $};
	\draw (343.67,317.07) node [anchor=north west][inner sep=0.75pt]  [color={rgb, 255:red, 208; green, 2; blue, 27 }  ,opacity=1 ]  {$\mathcal{B}_{2}$};
	\draw (415.33,189.07) node [anchor=north west][inner sep=0.75pt]  [font=\normalsize]  {$\skull $};
	\draw (242.33,67.07) node [anchor=north west][inner sep=0.75pt]  [color={rgb, 255:red, 74; green, 144; blue, 226 }  ,opacity=1 ]  {$\mathcal{B}_{2}^{\dagger }$};
	\draw (242.33,117.73) node [anchor=north west][inner sep=0.75pt]  [color={rgb, 255:red, 74; green, 144; blue, 226 }  ,opacity=1 ]  {$\mathcal{B}_{2}^{\dagger }$};
	\draw (208.33,125.07) node [anchor=north west][inner sep=0.75pt]  [color={rgb, 255:red, 0; green, 0; blue, 0 }  ,opacity=1 ]  {$\mathcal{B}_{2}^{\dagger }$};
	\draw (243,266.4) node [anchor=north west][inner sep=0.75pt]  [color={rgb, 255:red, 74; green, 144; blue, 226 }  ,opacity=1 ]  {$\mathcal{B}_{2}^{\dagger }$};
	\draw (243,316.4) node [anchor=north west][inner sep=0.75pt]  [color={rgb, 255:red, 74; green, 144; blue, 226 }  ,opacity=1 ]  {$\mathcal{B}_{2}^{\dagger }$};
	\draw (346.33,218.4) node [anchor=north west][inner sep=0.75pt]  [color={rgb, 255:red, 208; green, 2; blue, 27 }  ,opacity=1 ]  {$\mathcal{B}_{2}$};
	\draw (241.67,219.07) node [anchor=north west][inner sep=0.75pt]  [color={rgb, 255:red, 74; green, 144; blue, 226 }  ,opacity=1 ]  {$\mathcal{B}_{2}^{\dagger }$};
	\draw (290.67,12.73) node [anchor=north west][inner sep=0.75pt]    {$\mathcal{K}_{2}^{+}$};
	\draw (296,365.4) node [anchor=north west][inner sep=0.75pt]    {$\mathcal{K}_{2}^{-}$};
		\draw (372,91.4) node [anchor=north west][inner sep=0.75pt]    {$\{c_{n,p}\}$};
	
\end{tikzpicture}
}

	\end{center}
	\caption{(a) Schematic representation of the action of operators $\mathcal{A}_{2},\mathcal{A}^\dagger_{2}$  on the vectors $c_{n,p}$ (b) Schematic representation of the action of operators $\mathcal{B}_{2},\mathcal{B}^\dagger_{2}$ on the vectors $c_{n,p}$.}
	\label{fig2}
\end{figure}

In this case we consider $\Kc_2^+=\oplus_{n=1}^\infty\Hil_2(n)$, and $\Kc_2^-=\oplus_{n=-\infty}^{0}\Hil_2(n)$. Hence $\Hil_2=\Kc_2^+\oplus\Kc_2^-$ and it follows that, taken $f\in\Kc_2^+$ and $g\in\Kc_2^-$, $\langle f, g\rangle_2=0$.  These two spaces are {\em disconnected}, as $\Hil_2^+$ and $\Hil_2^-$ before: if we act on some $f\in\Kc_2^+$ with $\Bc_2$ or with $\Bc_2^\dagger$, and their powers, we get some other vector which is still in $\Kc_2^+$. The same if we start from a vector $g\in\Kc_2^-$. In other words, this is just a different choice with respect to the previous one, the main difference consisting in how we decompose the space $\Hil_2$. Of course, other possibilities are also possible, but we will restrict to the two described here.   
In this case our states in (\ref{cs8}) and (\ref{cs9}) should be replaced by the following ones:
\be
\Phi_\Bc^+(z_1,z_2)=e^{-(|z_1|^2+|z_2|^2)/2}\sum_{n1=0}^\infty\sum_{n2=0}^\infty\frac{z_1^{n_1}\,z_2^{n_2}}{\sqrt{n_1!\,n_2!}}\,c_{n_1,n_2+1},
\label{cs20}\en
and
\be
\Phi_\Bc^-(z_1,z_2)=e^{-(|z_1|^2+|z_2|^2)/2}\sum_{n1=0}^\infty\sum_{n2=0}^\infty\frac{z_1^{n_1}\,z_2^{n_2}}{\sqrt{n_1!\,n_2!}}\,c_{n_1,-n_2},
\label{cs21}\en
which clearly live respectively in $\Kc_2^+$ and $\Kc_2^-$. These states are normalized, mutually orthogonal, satisfy a resolution of the identity in $\Kc_2^+$ and $\Kc_2^-$, respectively, and are eigenstates of the annihilation operators in the following sense:
\be
\Ac_1\Phi_\Bc^+(z_1,z_2)=z_1\,\Phi_\Bc^+(z_1,z_2), \qquad \Ac_1\Phi_\Bc^-(z_1,z_2)=z_1\,\Phi_\Bc^-(z_1,z_2),
\label{cs22}\en
similar to (\ref{cs11}), and
\be
\Bc_2^\dagger\Phi_\Bc^+(z_1,z_2)=z_2\,\Phi_\Bc^+(z_1,z_2), \qquad \Bc_2\Phi_\Bc^-(z_1,z_2)=z_2\,\Phi_\Bc^-(z_1,z_2),
\label{cs23}\en
Our consideration for $\Phi_\Ac^\pm(z_1,z_2)$ could be repeated now, with minor changes, for $\Phi_\Bc^\pm(z_1,z_2)$.

%
%
%

\subsection{Working with tensor products}\label{subsect23}

What we have done in the first part of this section, up to Section \ref{subsect21} included, can be restated in terms of tensor product Hilbert spaces. This approach will be particularly useful in presence of a chemical potential, as we will see in Section \ref{sect3}. In view of this relevance, we briefly sketch here this alternative settings since it is easier, due to the fact that, as we will see later on, we are dealing with ONB, rather than with biorthogonal sets.

The fact that $n_1$ in (\ref{08})-(\ref{eigv}) plays essentially no role suggests to rewrite the o.n. vectors $e_{n_1,n_2}$ introduced after (\ref{09}) as $e_{n_1,n_2}=e_{n_1}^{(1)}\otimes e_{n_2}^{(2)}$, where $e_{n_j}^{(j)}=\frac{1}{\sqrt{n_j!}}\,(A_j^\dagger)^{n_j}e_0^{(j)}$, and $A_j e_0^{(j)}=0$, $j=1,2$. Hence we put
$$
v_{n_1,0}=e_{n_1}^{(1)}\otimes\begin{pmatrix}
	e_{n_2}^{(2)} \\
	0
\end{pmatrix} =e_{n_1}^{(1)}\otimes v_0
$$
and, for $n_2 \geq 1$,
$$
v_{n_1,n_2}^{\pm}= e_{n_1}^{(1)}\otimes\dfrac{1}{\sqrt{2}} \begin{pmatrix}
	e_{n_2}^{(2)} \\
	\mp i e_{n_2-1}^{(2)}
\end{pmatrix}= e_{n_1}^{(1)}\otimes v_{n_2}^\pm.
$$
Here we call $\Hil^{(j)}$ the Hilbert space spanned by $\F_j=\{e_{n}^{(j)}, \,n\geq0\}$, with scalar product $\langle.,.\rangle_{(j)}$, $j=1,2$, and $\Kc$ the Hilbert space spanned by $\F_v=\{v_n^\pm, \, n\geq1\}\cup\{v_0\}$, with scalar product $\langle f,g\rangle_{\Kc}=\langle f_1,g_1\rangle_2+\langle f_2,g_2\rangle_2$, for all $f=\begin{pmatrix}
	f_1 \\
	f_2
\end{pmatrix}$ and $g=\begin{pmatrix}
g_1 \\
g_2
\end{pmatrix}$, $f_j,g_j\in\Hil^{(2)}$, so that $f,g\in\Kc$. $\F_v$ is an ONB for $\Kc$, and $\tilde\F_v=\{v_{n_1,n_2}^\pm, \, n_1\geq0, n_2\geq1\}\cup\{v_{n_1,0},\,n_1\geq0\}$ is an ONB in $\tilde\Hil=\Hil^{(1)}\otimes\Kc$, which is an Hilbert space with respect to the scalar product
$$
\langle \tilde f,\tilde g\rangle_{\tilde\Hil}= \langle f^{(1)},g^{(1)}\rangle_{(1)} \langle f_\Kc,g_\Kc\rangle_{\Kc},
$$
where $\tilde f=f^{(1)}\otimes f_\Kc$ and $\tilde g=g^{(1)}\otimes g_\Kc$, with $f^{(1)},g^{(1)}\in\Hil^{(1)}$ and $f_\Kc,g_\Kc\in\Kc$.

As in (\ref{f1}) it is convenient to rewrite the vectors in $\tilde\F_v$ as follows: $\tilde\F_v=\{w_{n,p}, \, n\geq0, \, p\in\mathbb{Z}\}$, where 
\be
w_{n,p}=e_n^{(1)}\otimes v_p^{(2)}=e_n^{(1)}\otimes \left\{
\begin{array}{ll}
	v_{p}^{
		+}, \hspace{2cm} p\geq1,\\
	v_{0}, \hspace{2cm} p=0,\\
	v_{-p}^-,\hspace{1.8cm} p\leq-1,
\end{array}
\right.
\label{a1}\en
and where $n\geq0$.
Hence we have $\langle w_{n,p},w_{m,q}\rangle_{\tilde\Hil}=\delta_{n,m}\delta_{p,q}$, $\forall\,n,m\geq0$ and $\forall\,p,q\in\mathbb{Z}$. Moreover, if we put $H_0=\1^{(1)}\otimes H_K^+$, where $\1^{(1)}$ is the identity operator on $\Hil^{(1)}$, we have
\be
H_0w_{n,p}=\E_{n,p}c_{n,p},
\label{a2}\en
where $n\geq0$ while $p\in\mathbb{Z}$, and $\E_{n,p}$ is given in (\ref{f2}). Now it is an easy task to identify the ladder operators acting on $\Hil^{(1)}$ and, with some difference, on $\Kc$. We first consider $\A_{(1)}=A_{(1)}\otimes\1_\Kc$ and its adjoint $\A_{(1)}^\dagger=A_{(1)}^\dagger\otimes\1_\Kc$. Here $A_{(1)}$ is exactly $A_1$ in (\ref{03}). We adopt this new notation, since it is more useful here and in the following. Also, we notice that we are using the same symbol for the adjoint in $\Hil^{(1)}$ and in $\tilde \Hil$. In fact, we will use the same symbol for all the adjoints we will introduce, here and in the following, since no confusion can arise. It is simple to check that
\be
\A_{(1)}\,w_{n,p}= \left\{
\begin{array}{ll}
	0, \hspace{1.8cm} n=0,\\
	\sqrt{n}w_{n-1,p}\hspace{0.7cm} n\geq1,
\end{array}
\right. \qquad\qquad \A_{(1)}^\dagger w_{n,p}=\sqrt{n+1}\,w_{n+1,p}, \quad \forall n\geq0,
\label{a3}\en 
 $p\in\mathbb{Z}$. We can use the following formula for $A_{(1)}$ and $\A_{(1)}$:
 \be
 A_{(1)}=\sum_{n=0}^\infty\sqrt{n+1}\,|e_n^{(1)}\rangle \langle e_{n+1}^{(1)}|, \qquad \A_{(1)}=\left(\sum_{n=0}^\infty\sqrt{n+1}\,|e_n^{(1)}\rangle \langle e_{n+1}^{(1)}|\right)\otimes \1_\Kc.
 \label{a4}\en
 These two formulas, and our previous discussion, suggest us to introduce further the operator $\A_\Kc$ as follows:
 \be
 A_\Kc=\sum_{p=-\infty}^\infty\sqrt{|p+1|}\,|v_p^{(2)}\rangle \langle v_{p+1}^{(2)}|, \qquad \A_\Kc=\1^{(1)}\otimes\left(\sum_{p=-\infty}^\infty\sqrt{|p+1|}\,|v_p^{(2)}\rangle \langle v_{p+1}^{(2)}|\right),
 \label{a5}\en
 i.e., $\A_\Kc=\1^{(1)}\otimes A_\Kc$. It is easy to check that
 \be
 \A_\Kc\,w_{n,p}=\sqrt{|p|}\, w_{n,p-1}, \qquad \qquad \A_\Kc^\dagger\,w_{n,p}=\sqrt{|p+1|}\, w_{n,p+1},
 \label{a6}\en
 for all $n\geq0$ and $p\in\mathbb{Z}$. $\A_\Kc^\dagger$ can be written as
 \be
 \A_\Kc^\dagger=\1^{(1)}\otimes\left(\sum_{p=-\infty}^\infty\sqrt{|p+1|}\,|v_{p+1}^{(2)}\rangle \langle v_{p}^{(2)}|\right).
 \label{a7}\en
 
We could, of course, adopt different definitions of these operators. In particular, we could repeat here what we have considered in Section \ref{sect22}, but since this second decomposition will not be used in the following (in fact, it was only meant to give a flavor of the freedom we have, in our construction!), we will not do that here. As for the coherent states in Section \ref{subsect21}, they could be restated again in terms of tensor products. But we postpone this possibility to the more interesting case discussed in the next section, i.e., to the case where a non zero chemical potential is introduced, as in \cite{BagaHatano_2016}.

\section{Chemical potential: $0<V<1$} \label{sect3}

In \cite{BagaHatano_2016} the authors considered a  slightly extended version of $H_K^+$ in (\ref{06}):
\be \label{31}
H(V) = \dfrac{2 i v_F}{\xi}   \begin{pmatrix}
	V & {A_2}^\dagger \\
	-{A_2} & -V
\end{pmatrix},  
\en 
where $V$ is a fixed positive quantity. Its physical meaning as a chemical potential is discussed in some details in \cite{BagaHatano_2016}. One of the crucial difference between $H_K^+$ and $H(V)$ is that this latter is not self-adjoint, if $V\neq0$. Infact $H^\dagger(V)=H(-V)$. In what follows we will analyze some aspects of $H(V)$, restricting first to the case  $V\in[0,1[$, and considering the case $V>1$ later on.

It is now a long, but simple, exercise to extend what we have done here in Section \ref{subsect23} to rewrite some of the results in \cite{BagaHatano_2016} in the following, more convenient, form.
\be
\varphi_{n_1,n_2}^\pm=e_{n_1}^{(1)}\otimes \varphi_{n_2}^{(\pm)}, \qquad \varphi_{n_1,0}=e_{n_1}^{(1)}\otimes \varphi_{0},
\label{32}\en
with $n_1,n_2\geq0$ and where
\be
\varphi_{0}=\begin{pmatrix}
	e_{0}^{(2)} \\
	0
\end{pmatrix}, \qquad
 \varphi_{n_2}^\pm= K_{n_2}^{\pm}(\varphi)\begin{pmatrix}
 	e_{n_2}^{(2)} \\
 	\alpha_{n_2}^{\pm}e_{n_2-1}^{(2)}
 \end{pmatrix}, \quad \mbox{if } n_2\geq1. 
\label{33} \en
Here
\be
\alpha_{n_2}^{\pm}=\frac{-V\mp i\sqrt{n_2-V^2}}{\sqrt{n_2}},
\label{34}\en
while $K_{n_2}^{\pm}(\varphi)$ is a normalization constant which will be fixed later. Notice that, since $0\leq V<1$, $\sqrt{n_2-V^2}$ is always real for each $n_2\geq1$. Hence $\alpha_{n_2}^{\pm}$ is never real.

We now further introduce the vectors
\be
\psi_{n_1,n_2}^\pm=e_{n_1}^{(1)}\otimes \psi_{n_2}^{(\pm)}, \qquad \psi_{n_1,0}=e_{n_1}^{(1)}\otimes \psi_{0},
\label{35}\en
with $n_1,n_2\geq0$ and where
\be
\psi_{0}=\begin{pmatrix}
	e_{0}^{(2)} \\
	0
\end{pmatrix}=\varphi_{0}, \qquad
\psi_{n_2}^\pm= K_{n_2}^{\pm}(\psi)\begin{pmatrix}
	e_{n_2}^{(2)} \\
	-\alpha_{n_2}^{\mp}e_{n_2-1}^{(2)}
\end{pmatrix}, \quad \mbox{if } n_2\geq1. 
\label{36} \en
As we did in (\ref{f1}) it is more convenient to rename these vectors as follows:
\be
x_{n,p}=e_n^{(1)}\otimes\varphi_p, \qquad y_{n,p}=e_n^{(1)}\otimes\psi_p,
\label{37}\en
where $n\geq0$, $p\in\mathbb{Z}$, and where we have introduced, in analogy with (\ref{a1}),
\be
\varphi_p=\left\{
\begin{array}{ll}
	\varphi_p^{
		+}, \hspace{1cm} p\geq1,\\
	\varphi_{0}, \hspace{1cm} p=0,\\
	\varphi_{-p}^-,\hspace{0.8cm} p\leq-1,
\end{array}
\right. \quad \mbox{and } \quad \psi_p=\left\{
\begin{array}{ll}
	\psi_p^{
		+}, \hspace{1cm} p\geq1,\\
	\psi_{0}, \hspace{1cm} p=0,\\
	\psi_{-p}^-,\hspace{0.8cm} p\leq-1.
\end{array}
\right.
\label{38}\en
If we now assume that
\be
\overline{K_{p}^{\pm}(\varphi)}\,K_{p}^{\pm}(\psi)=\frac{p}{2(p-V^2\pm iV\sqrt{p-V^2})},
\label{39}\en
$p\geq1$, we conclude that
\be
\langle x_{n,p},y_{m,q}\rangle_{\tilde\Hil}= \langle e_n^{(1)},e_m^{(1)}\rangle_{(1)} \langle\varphi_p,\psi_q\rangle_{\Kc}=\delta_{n,m}\delta_{p,q},
\label{310}\en
$n,m\geq0$ and $p,q\in\mathbb{Z}$. This means that the two sets $\F_x=\{x_{n,p}\}$ and $\F_y=\{y_{n,q}\}$ are biorthonormal in $\tilde\Hil$. It is also possible to check that any vector $\tilde f\in\tilde\Hil$, $\tilde f=f^{(1)}\otimes f_\Kc$ with $f^{(1)}\in\Hil^{(1)}$ and $f_\Kc\in\Kc$, which is orthogonal to all the $x_{n,p}$, or to all the $y_{n,p}$, is necessarily zero. This is because the sets $\{\varphi_p\}$ and $\{\psi_p\}$ are both total in $\Kc$, as it is easy to check.

Now, if we put $\tilde H(V)=\1^{(1)}\otimes H(V)$, we also have  $\tilde H^\dagger(V)=\1^{(1)}\otimes H(-V)$ and
\be
\left\{
\begin{array}{ll}
	\tilde H(V)\,x_{n,p}=\E_{p}\,x_{n,p},\\
	\tilde H^\dagger(V)\,y_{n,p}=\overline{\E_{p}}\,y_{n,p},
\end{array}
\right.
\label{311}\en
with
\be
\E_{p}=\left\{
\begin{array}{ll}
	\frac{2v_F}{\xi}\,\sqrt{p-V^2} \hspace{2.6cm}p\geq1,\\
	\frac{2iv_F}{\xi}\,V, \hspace{3.6cm} p=0,\\
		-\frac{2v_F}{\xi}\,\sqrt{-p-V^2},\hspace{1.7cm} p\leq-1.
\end{array}
\right.
\label{312}\en
With our choice of $V$, all the eigenvalues of $\tilde H(V)$ and its adjoint are real, except one, $\E_0$. This is important, since it implies that $\tilde H(V)$ and $\tilde H^\dagger(V)$ are not isospectral, at least if $V\neq0$. Hence it is not possible, in principle, to find an operator which intertwines between the two, \cite{IntOper_2005}. Also, see \cite{BagaHatano_2021}, it is not possible to define  on $\tilde \Hil$ any  scalar product which makes of $\tilde H(V)$ a self-adjoint operator (with respect to the adjoint map defined by this new scalar product). This is never possible whenever the Hamiltonian operator has, at least, one complex eigenvalue.

As in (\ref{a4}) we introduce the operators $\A_{(1)}$ and $\A_{(1)}^\dagger$, since they act as ladder operators also on the families $\F_x$ and $\F_y$:
\be
\A_{(1)}\,x_{n,p}= \left\{
\begin{array}{ll}
	0, \hspace{1.8cm} n=0,\\
	\sqrt{n}x_{n-1,p}\hspace{0.7cm} n\geq1,
\end{array}
\right. \qquad\qquad \A_{(1)}^\dagger x_{n,p}=\sqrt{n+1}\,x_{n+1,p}, \quad \forall n\geq0,
\label{312b}\en 
as well as
\be
\A_{(1)}\,y_{n,p}= \left\{
\begin{array}{ll}
	0, \hspace{1.8cm} n=0,\\
	\sqrt{n}y_{n-1,p}\hspace{0.7cm} n\geq1,
\end{array}
\right. \qquad\qquad \A_{(1)}^\dagger y_{n,p}=\sqrt{n+1}\,y_{n+1,p}, \quad \forall n\geq0,
\label{312c}\en 
$\forall\,p\in\mathbb{Z}$.
 The operator $\A_\Kc$ in (\ref{a5}) must be replaced by the following 
 \be
A_\Kc(V)=\sum_{p=-\infty}^\infty\sqrt{|p+1|}\,|\varphi_p\rangle \langle \psi_{p+1}|, \qquad \A_\Kc(V)=\1^{(1)}\otimes\left(\sum_{p=-\infty}^\infty\sqrt{|p+1|}\,|\varphi_p\rangle \langle \psi_{p+1}|\right),
\label{313}\en
so that $\A_\Kc(V)=\1^{(1)}\otimes A_\Kc(V)$. Then we have
\be
\A_\Kc(V)\,x_{n,p}=\sqrt{|p|}\, x_{n,p-1}, \qquad \qquad \A_\Kc^\dagger(V)\,y_{n,p}=\sqrt{|p+1|}\, y_{n,p+1},
\label{314}\en
for all $n\geq0$ and $p\in\mathbb{Z}$. In particular we see that $\A_\Kc(V)\,x_{n,0}=\A_\Kc^\dagger(V)\,y_{n,-1}=0$, for all $n\geq0$.   The proof of all these equations are standard, and will not be given here. We only observe that all these operators are defined on sets which are total in $\tilde \Hil$. For instance, $D(\A_\Kc(V))\supseteq\Lc_x$, the linear span of the $x_{n,p}$'s. The fact that $\F_x$ is total can be deduced as in \cite{BagaHatano_2016}. 

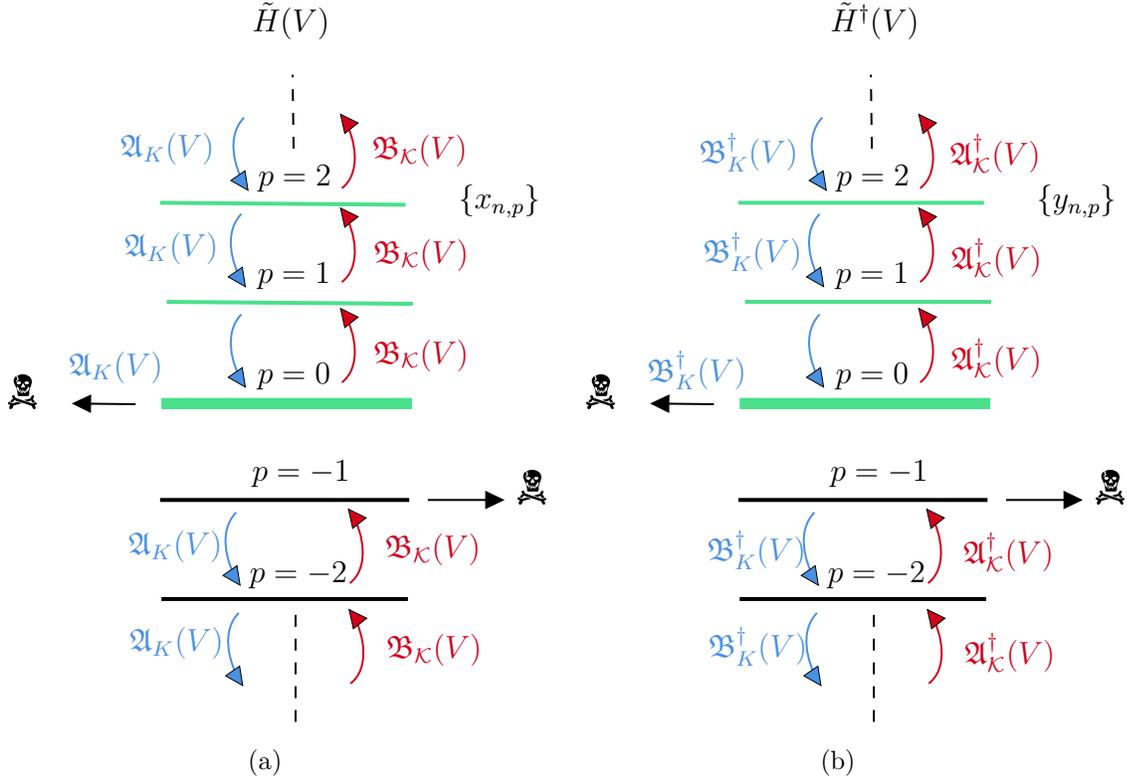
\begin{figure}[!ht]

\subfigure[]{\tikzset{every picture/.style={line width=0.75pt}} 

\begin{tikzpicture}[x=0.75pt,y=0.75pt,yscale=-1,xscale=1]
	
	\draw [color={rgb, 255:red, 74; green, 224; blue, 140 }  ,draw opacity=1 ][line width=1.5]    (239,188.67) -- (363.17,189.67) ;
	\draw [color={rgb, 255:red, 74; green, 224; blue, 140 }  ,draw opacity=1 ][line width=1.5]    (235.67,138.67) -- (359.5,139.67) ;
	\draw [color={rgb, 255:red, 208; green, 2; blue, 27 }  ,draw opacity=1 ]   (327.33,178.67) .. controls (337.15,171.31) and (334.55,153.79) .. (328.36,143.77) ;
	\draw [shift={(326.67,141.33)}, rotate = 51.84] [fill={rgb, 255:red, 208; green, 2; blue, 27 }  ,fill opacity=1 ][line width=0.08]  [draw opacity=0] (8.93,-4.29) -- (0,0) -- (8.93,4.29) -- cycle    ;
	\draw [color={rgb, 255:red, 208; green, 2; blue, 27 }  ,draw opacity=1 ]   (327.33,131.33) .. controls (337.15,123.97) and (334.55,106.46) .. (328.36,96.43) ;
	\draw [shift={(326.67,94)}, rotate = 51.84] [fill={rgb, 255:red, 208; green, 2; blue, 27 }  ,fill opacity=1 ][line width=0.08]  [draw opacity=0] (8.93,-4.29) -- (0,0) -- (8.93,4.29) -- cycle    ;
	\draw [color={rgb, 255:red, 74; green, 144; blue, 226 }  ,draw opacity=1 ]   (278.39,129.79) .. controls (273.05,120.55) and (268.9,105.07) .. (278,95.67) ;
	\draw [shift={(280,132.33)}, rotate = 235.3] [fill={rgb, 255:red, 74; green, 144; blue, 226 }  ,fill opacity=1 ][line width=0.08]  [draw opacity=0] (8.93,-4.29) -- (0,0) -- (8.93,4.29) -- cycle    ;
	\draw [color={rgb, 255:red, 74; green, 144; blue, 226 }  ,draw opacity=1 ]   (277.06,178.45) .. controls (271.72,169.21) and (267.57,153.74) .. (276.67,144.33) ;
	\draw [shift={(278.67,181)}, rotate = 235.3] [fill={rgb, 255:red, 74; green, 144; blue, 226 }  ,fill opacity=1 ][line width=0.08]  [draw opacity=0] (8.93,-4.29) -- (0,0) -- (8.93,4.29) -- cycle    ;
	\draw [color={rgb, 255:red, 74; green, 224; blue, 140 }  ,draw opacity=1 ][line width=4.25]    (236.33,240) -- (362.67,240) ;
	\draw [color={rgb, 255:red, 0; green, 0; blue, 0 }  ,draw opacity=1 ][line width=1.5]    (236.33,338.67) -- (360.67,338.67) ;
	\draw [color={rgb, 255:red, 0; green, 0; blue, 0 }  ,draw opacity=1 ][line width=1.5]    (235.67,288.67) -- (361.33,288.67) ;
	\draw [color={rgb, 255:red, 208; green, 2; blue, 27 }  ,draw opacity=1 ]   (331.33,330.67) .. controls (341.15,323.31) and (338.55,305.79) .. (332.36,295.77) ;
	\draw [shift={(330.67,293.33)}, rotate = 51.84] [fill={rgb, 255:red, 208; green, 2; blue, 27 }  ,fill opacity=1 ][line width=0.08]  [draw opacity=0] (8.93,-4.29) -- (0,0) -- (8.93,4.29) -- cycle    ;
	\draw  [dash pattern={on 4.5pt off 4.5pt}]  (303,111.33) -- (302.5,74) ;
	\draw [color={rgb, 255:red, 74; green, 144; blue, 226 }  ,draw opacity=1 ]   (274.39,329.12) .. controls (269.05,319.88) and (264.9,304.4) .. (274,295) ;
	\draw [shift={(276,331.67)}, rotate = 235.3] [fill={rgb, 255:red, 74; green, 144; blue, 226 }  ,fill opacity=1 ][line width=0.08]  [draw opacity=0] (8.93,-4.29) -- (0,0) -- (8.93,4.29) -- cycle    ;
	\draw  [dash pattern={on 4.5pt off 4.5pt}]  (304,400.67) -- (304,343.33) ;
	\draw    (223.33,240.33) -- (194.67,240.03) ;
	\draw [shift={(191.67,240)}, rotate = 0.6] [fill={rgb, 255:red, 0; green, 0; blue, 0 }  ][line width=0.08]  [draw opacity=0] (8.93,-4.29) -- (0,0) -- (8.93,4.29) -- cycle    ;
	\draw [color={rgb, 255:red, 208; green, 2; blue, 27 }  ,draw opacity=1 ]   (331.67,381.33) .. controls (341.48,373.97) and (338.88,356.46) .. (332.69,346.43) ;
	\draw [shift={(331,344)}, rotate = 51.84] [fill={rgb, 255:red, 208; green, 2; blue, 27 }  ,fill opacity=1 ][line width=0.08]  [draw opacity=0] (8.93,-4.29) -- (0,0) -- (8.93,4.29) -- cycle    ;
	\draw [color={rgb, 255:red, 74; green, 144; blue, 226 }  ,draw opacity=1 ]   (275.72,379.79) .. controls (270.38,370.55) and (266.23,355.07) .. (275.33,345.67) ;
	\draw [shift={(277.33,382.33)}, rotate = 235.3] [fill={rgb, 255:red, 74; green, 144; blue, 226 }  ,fill opacity=1 ][line width=0.08]  [draw opacity=0] (8.93,-4.29) -- (0,0) -- (8.93,4.29) -- cycle    ;
	\draw    (370.67,288.67) -- (405.83,288.67) ;
	\draw [shift={(408.83,288.67)}, rotate = 180] [fill={rgb, 255:red, 0; green, 0; blue, 0 }  ][line width=0.08]  [draw opacity=0] (8.93,-4.29) -- (0,0) -- (8.93,4.29) -- cycle    ;
	\draw [color={rgb, 255:red, 208; green, 2; blue, 27 }  ,draw opacity=1 ]   (327.33,229.33) .. controls (337.15,221.97) and (334.55,204.46) .. (328.36,194.43) ;
	\draw [shift={(326.67,192)}, rotate = 51.84] [fill={rgb, 255:red, 208; green, 2; blue, 27 }  ,fill opacity=1 ][line width=0.08]  [draw opacity=0] (8.93,-4.29) -- (0,0) -- (8.93,4.29) -- cycle    ;
	\draw [color={rgb, 255:red, 74; green, 144; blue, 226 }  ,draw opacity=1 ]   (277.06,229.12) .. controls (271.72,219.88) and (267.57,204.4) .. (276.67,195) ;
	\draw [shift={(278.67,231.67)}, rotate = 235.3] [fill={rgb, 255:red, 74; green, 144; blue, 226 }  ,fill opacity=1 ][line width=0.08]  [draw opacity=0] (8.93,-4.29) -- (0,0) -- (8.93,4.29) -- cycle    ;
	
	\draw (283.33,167.4) node [anchor=north west][inner sep=0.75pt]    {$p=1$};
	\draw (283.67,118.07) node [anchor=north west][inner sep=0.75pt]    {$p=2$};
	\draw (341,102.73) node [anchor=north west][inner sep=0.75pt]  [color={rgb, 255:red, 208; green, 2; blue, 27 }  ,opacity=1 ]  {$\mathfrak{B}_{\mathcal{K}}(V)$};
	\draw (279.33,318.4) node [anchor=north west][inner sep=0.75pt]    {$p=-2$};
	\draw (156.33,222.4) node [anchor=north west][inner sep=0.75pt]  [font=\normalsize]  {$\skull $};
	\draw (413.33,270.73) node [anchor=north west][inner sep=0.75pt]  [font=\normalsize]  {$\skull $};
	\draw (214.33,102.07) node [anchor=north west][inner sep=0.75pt]  [color={rgb, 255:red, 74; green, 144; blue, 226 }  ,opacity=1 ]  {$\mathfrak{A}_{K}( V)$};
	\draw (384.67,128.07) node [anchor=north west][inner sep=0.75pt]    {$\{x_{n,p}\}$};
	\draw (280.67,36.07) node [anchor=north west][inner sep=0.75pt]    {$\tilde{H}( V)$};
	\draw (216.33,152.07) node [anchor=north west][inner sep=0.75pt]  [color={rgb, 255:red, 74; green, 144; blue, 226 }  ,opacity=1 ]  {$\mathfrak{A}_{K}( V)$};
	\draw (341.67,155.4) node [anchor=north west][inner sep=0.75pt]  [color={rgb, 255:red, 208; green, 2; blue, 27 }  ,opacity=1 ]  {$\mathfrak{B}_{\mathcal{K}}(V)$};
	\draw (283.33,218.07) node [anchor=north west][inner sep=0.75pt]    {$p=0$};
	\draw (341.67,206.07) node [anchor=north west][inner sep=0.75pt]  [color={rgb, 255:red, 208; green, 2; blue, 27 }  ,opacity=1 ]  {$\mathfrak{B}_{\mathcal{K}}(V)$};
	\draw (188.33,212.07) node [anchor=north west][inner sep=0.75pt]  [color={rgb, 255:red, 74; green, 144; blue, 226 }  ,opacity=1 ]  {$\mathfrak{A}_{K}( V)$};
	\draw (218.33,302.73) node [anchor=north west][inner sep=0.75pt]  [color={rgb, 255:red, 74; green, 144; blue, 226 }  ,opacity=1 ]  {$\mathfrak{A}_{K}( V)$};
	\draw (219,351.4) node [anchor=north west][inner sep=0.75pt]  [color={rgb, 255:red, 74; green, 144; blue, 226 }  ,opacity=1 ]  {$\mathfrak{A}_{K}( V)$};
	\draw (347.67,304.07) node [anchor=north west][inner sep=0.75pt]  [color={rgb, 255:red, 208; green, 2; blue, 27 }  ,opacity=1 ]  {$\mathfrak{B}_{\mathcal{K}}(V)$};
	\draw (348.33,354.73) node [anchor=north west][inner sep=0.75pt]  [color={rgb, 255:red, 208; green, 2; blue, 27 }  ,opacity=1 ]  {$\mathfrak{B}_{\mathcal{K}}(V)$};
	\draw (280.67,267.07) node [anchor=north west][inner sep=0.75pt]    {$p=-1$};

\end{tikzpicture}}
\subfigure[]{
\tikzset{every picture/.style={line width=0.75pt}} 

\begin{tikzpicture}[x=0.75pt,y=0.75pt,yscale=-1,xscale=1]
	
	\draw [color={rgb, 255:red, 74; green, 224; blue, 140 }  ,draw opacity=1 ][line width=1.5]    (239,188.67) -- (363.17,188.67) ;
	\draw [color={rgb, 255:red, 74; green, 224; blue, 140 }  ,draw opacity=1 ][line width=1.5]    (235.67,138.67) -- (359.5,138.67) ;
	\draw [color={rgb, 255:red, 208; green, 2; blue, 27 }  ,draw opacity=1 ]   (327.33,178.67) .. controls (337.15,171.31) and (334.55,153.79) .. (328.36,143.77) ;
	\draw [shift={(326.67,141.33)}, rotate = 51.84] [fill={rgb, 255:red, 208; green, 2; blue, 27 }  ,fill opacity=1 ][line width=0.08]  [draw opacity=0] (8.93,-4.29) -- (0,0) -- (8.93,4.29) -- cycle    ;
	\draw [color={rgb, 255:red, 208; green, 2; blue, 27 }  ,draw opacity=1 ]   (327.33,131.33) .. controls (337.15,123.97) and (334.55,106.46) .. (328.36,96.43) ;
	\draw [shift={(326.67,94)}, rotate = 51.84] [fill={rgb, 255:red, 208; green, 2; blue, 27 }  ,fill opacity=1 ][line width=0.08]  [draw opacity=0] (8.93,-4.29) -- (0,0) -- (8.93,4.29) -- cycle    ;
	\draw [color={rgb, 255:red, 74; green, 144; blue, 226 }  ,draw opacity=1 ]   (278.39,129.79) .. controls (273.05,120.55) and (268.9,105.07) .. (278,95.67) ;
	\draw [shift={(280,132.33)}, rotate = 235.3] [fill={rgb, 255:red, 74; green, 144; blue, 226 }  ,fill opacity=1 ][line width=0.08]  [draw opacity=0] (8.93,-4.29) -- (0,0) -- (8.93,4.29) -- cycle    ;
	\draw [color={rgb, 255:red, 74; green, 144; blue, 226 }  ,draw opacity=1 ]   (277.06,178.45) .. controls (271.72,169.21) and (267.57,153.74) .. (276.67,144.33) ;
	\draw [shift={(278.67,181)}, rotate = 235.3] [fill={rgb, 255:red, 74; green, 144; blue, 226 }  ,fill opacity=1 ][line width=0.08]  [draw opacity=0] (8.93,-4.29) -- (0,0) -- (8.93,4.29) -- cycle    ;
	\draw [color={rgb, 255:red, 74; green, 224; blue, 140 }  ,draw opacity=1 ][line width=4.25]    (236.33,240) -- (362.67,240) ;
	\draw [color={rgb, 255:red, 0; green, 0; blue, 0 }  ,draw opacity=1 ][line width=1.5]    (236.33,338.67) -- (360.67,338.67) ;
	\draw [color={rgb, 255:red, 0; green, 0; blue, 0 }  ,draw opacity=1 ][line width=1.5]    (235.67,288.67) -- (361.33,288.67) ;
	\draw [color={rgb, 255:red, 208; green, 2; blue, 27 }  ,draw opacity=1 ]   (331.33,330.67) .. controls (341.15,323.31) and (338.55,305.79) .. (332.36,295.77) ;
	\draw [shift={(330.67,293.33)}, rotate = 51.84] [fill={rgb, 255:red, 208; green, 2; blue, 27 }  ,fill opacity=1 ][line width=0.08]  [draw opacity=0] (8.93,-4.29) -- (0,0) -- (8.93,4.29) -- cycle    ;
	\draw  [dash pattern={on 4.5pt off 4.5pt}]  (303,111.33) -- (302.5,74) ;
	\draw [color={rgb, 255:red, 74; green, 144; blue, 226 }  ,draw opacity=1 ]   (274.39,329.12) .. controls (269.05,319.88) and (264.9,304.4) .. (274,295) ;
	\draw [shift={(276,331.67)}, rotate = 235.3] [fill={rgb, 255:red, 74; green, 144; blue, 226 }  ,fill opacity=1 ][line width=0.08]  [draw opacity=0] (8.93,-4.29) -- (0,0) -- (8.93,4.29) -- cycle    ;
	\draw  [dash pattern={on 4.5pt off 4.5pt}]  (304,400.67) -- (304,343.33) ;
	\draw    (223.33,240.33) -- (194.67,240.03) ;
	\draw [shift={(191.67,240)}, rotate = 0.6] [fill={rgb, 255:red, 0; green, 0; blue, 0 }  ][line width=0.08]  [draw opacity=0] (8.93,-4.29) -- (0,0) -- (8.93,4.29) -- cycle    ;
	\draw [color={rgb, 255:red, 208; green, 2; blue, 27 }  ,draw opacity=1 ]   (331.67,381.33) .. controls (341.48,373.97) and (338.88,356.46) .. (332.69,346.43) ;
	\draw [shift={(331,344)}, rotate = 51.84] [fill={rgb, 255:red, 208; green, 2; blue, 27 }  ,fill opacity=1 ][line width=0.08]  [draw opacity=0] (8.93,-4.29) -- (0,0) -- (8.93,4.29) -- cycle    ;
	\draw [color={rgb, 255:red, 74; green, 144; blue, 226 }  ,draw opacity=1 ]   (275.72,379.79) .. controls (270.38,370.55) and (266.23,355.07) .. (275.33,345.67) ;
	\draw [shift={(277.33,382.33)}, rotate = 235.3] [fill={rgb, 255:red, 74; green, 144; blue, 226 }  ,fill opacity=1 ][line width=0.08]  [draw opacity=0] (8.93,-4.29) -- (0,0) -- (8.93,4.29) -- cycle    ;
	\draw    (370.67,288.67) -- (405.83,288.67) ;
	\draw [shift={(408.83,288.67)}, rotate = 180] [fill={rgb, 255:red, 0; green, 0; blue, 0 }  ][line width=0.08]  [draw opacity=0] (8.93,-4.29) -- (0,0) -- (8.93,4.29) -- cycle    ;
	\draw [color={rgb, 255:red, 208; green, 2; blue, 27 }  ,draw opacity=1 ]   (327.33,229.33) .. controls (337.15,221.97) and (334.55,204.46) .. (328.36,194.43) ;
	\draw [shift={(326.67,192)}, rotate = 51.84] [fill={rgb, 255:red, 208; green, 2; blue, 27 }  ,fill opacity=1 ][line width=0.08]  [draw opacity=0] (8.93,-4.29) -- (0,0) -- (8.93,4.29) -- cycle    ;
	\draw [color={rgb, 255:red, 74; green, 144; blue, 226 }  ,draw opacity=1 ]   (277.06,229.12) .. controls (271.72,219.88) and (267.57,204.4) .. (276.67,195) ;
	\draw [shift={(278.67,231.67)}, rotate = 235.3] [fill={rgb, 255:red, 74; green, 144; blue, 226 }  ,fill opacity=1 ][line width=0.08]  [draw opacity=0] (8.93,-4.29) -- (0,0) -- (8.93,4.29) -- cycle    ;
	
	\draw (283.33,167.4) node [anchor=north west][inner sep=0.75pt]    {$p=1$};
	\draw (283.67,118.07) node [anchor=north west][inner sep=0.75pt]    {$p=2$};
	\draw (341,102.73) node [anchor=north west][inner sep=0.75pt]  [color={rgb, 255:red, 208; green, 2; blue, 27 }  ,opacity=1 ]  {$\mathfrak{A}^\dagger_{\mathcal{K}}(V)$};
	\draw (279.33,318.4) node [anchor=north west][inner sep=0.75pt]    {$p=-2$};
	\draw (156.33,222.4) node [anchor=north west][inner sep=0.75pt]  [font=\normalsize]  {$\skull $};
	\draw (413.33,270.73) node [anchor=north west][inner sep=0.75pt]  [font=\normalsize]  {$\skull $};
	\draw (214.33,102.07) node [anchor=north west][inner sep=0.75pt]  [color={rgb, 255:red, 74; green, 144; blue, 226 }  ,opacity=1 ]  {$\mathfrak{B}^\dagger_{K}( V)$};
	\draw (384.67,128.07) node [anchor=north west][inner sep=0.75pt]    {$\{y_{n,p}\}$};
	\draw (280.67,36.07) node [anchor=north west][inner sep=0.75pt]    {$\tilde{H}^\dagger(V)$};
	\draw (216.33,152.07) node [anchor=north west][inner sep=0.75pt]  [color={rgb, 255:red, 74; green, 144; blue, 226 }  ,opacity=1 ]  {$\mathfrak{B}^\dagger_{K}( V)$};
	\draw (341.67,155.4) node [anchor=north west][inner sep=0.75pt]  [color={rgb, 255:red, 208; green, 2; blue, 27 }  ,opacity=1 ]  {$\mathfrak{A}^\dagger_{\mathcal{K}}(V)$};
	\draw (283.33,218.07) node [anchor=north west][inner sep=0.75pt]    {$p=0$};
	\draw (341.67,206.07) node [anchor=north west][inner sep=0.75pt]  [color={rgb, 255:red, 208; green, 2; blue, 27 }  ,opacity=1 ]  {$\mathfrak{A}^\dagger_{\mathcal{K}}(V)$};
	\draw (188.33,212.07) node [anchor=north west][inner sep=0.75pt]  [color={rgb, 255:red, 74; green, 144; blue, 226 }  ,opacity=1 ]  {$\mathfrak{B}^\dagger_{K}( V)$};
	\draw (218.33,302.73) node [anchor=north west][inner sep=0.75pt]  [color={rgb, 255:red, 74; green, 144; blue, 226 }  ,opacity=1 ]  {$\mathfrak{B}^\dagger_{K}( V)$};
	\draw (219,351.4) node [anchor=north west][inner sep=0.75pt]  [color={rgb, 255:red, 74; green, 144; blue, 226 }  ,opacity=1 ]  {$\mathfrak{B}^\dagger_{K}( V)$};
	\draw (347.67,304.07) node [anchor=north west][inner sep=0.75pt]  [color={rgb, 255:red, 208; green, 2; blue, 27 }  ,opacity=1 ]  {$\mathfrak{A}^\dagger_{\mathcal{K}}(V)$};
	\draw (348.33,354.73) node [anchor=north west][inner sep=0.75pt]  [color={rgb, 255:red, 208; green, 2; blue, 27 }  ,opacity=1 ]  {$\mathfrak{A}^\dagger_{\mathcal{K}}(V)$};
	\draw (280.67,267.07) node [anchor=north west][inner sep=0.75pt]    {$p=-1$};

\end{tikzpicture}}
	\caption{(a) Schematic representation of the action of the operators $\mathfrak{A}_{\mathcal{K}}(V),\mathfrak{B}_{\mathcal{K}}(V)$ on the vectors $x_{n,p}$ (b) Schematic representation of the action of the operators $\mathfrak{A}^\dagger_{\mathcal{K}}(V),\mathfrak{B}^\dagger_{\mathcal{K}}(V)$ on the vectors $y_{n,p}$.}
	\label{fig3}
\end{figure}

It is interesting to notice that, as (\ref{314}) shows, $\A_\Kc(V)$ is a lowering operator for $\F_x$,  while $\A_\Kc^\dagger(V)$ is a raising operator for $\F_y$. This is a normal situation in presence of ladder operators and biorthonormal families, \cite{bagspringer}. What is still missing, and it is useful to have, is a raising operator for $\F_x$ and a lowering operator for $\F_y$. These operators can be easily defined:
\be
\B_\Kc(V)=\1^{(1)}\otimes\left(\sum_{p=-\infty}^\infty\sqrt{|p+1|}\,|\varphi_{p+1}\rangle \langle \psi_{p}|\right),\quad \B_\Kc^\dagger(V)=\1^{(1)}\otimes\left(\sum_{p=-\infty}^\infty\sqrt{|p+1|}\,|\psi_{p}\rangle \langle \varphi_{p+1}|\right).
\label{315}\en
Indeed we have that
\be
\B_\Kc(V)\,x_{n,p}=\sqrt{|p+1|}\, x_{n,p+1}, \qquad \qquad \B_\Kc^\dagger(V)\,y_{n,p}=\sqrt{|p|}\, y_{n,p-1},
\label{316}\en
for all $n\geq0$ and $p\in\mathbb{Z}$. This is exactly what we were looking for. Again, these operators are defined on rather large sets: for instance $D(\B_\Kc(V))\supseteq\Lc_x$ and $D(\B_\Kc^\dagger(V))\supseteq\Lc_y$. Here $\Lc_y$ is  the linear span of the  $y_{n,p}$'s. Figure \ref{fig2} is now replaced by the Figure \ref{fig3}, where we put together the energetic levels of  $\tilde H(V)$ and $\tilde H^\dagger(V)$. We recall (and this is clear from Figure \ref{fig3}), that all the energy levels (except the one with $p=0$) for $\tilde H(V)$ and $\tilde H^\dagger(V)$ coincides. We put in the figure also the energetic levels for $p=0$ which, however, are purely imaginary. For this reason we use a {bold line} in the figure. The fact that we put this zero level between $p=1$ and $p=-1$ does not mean at all, of course, that the energy of the zero level is also between these two energy levels. It is only because $0$ (and not $\E_0$) is between $-1$ and $+1$.

If we now proceed as in Section \ref{subsect21}, we introduce the vectors
\be
\varphi^+(z_1,z_2)=e^{-(|z_1|^2+|z_2|^2)/2}\sum_{n1=0}^\infty\sum_{n2=0}^\infty\frac{z_1^{n_1}\,z_2^{n_2}}{\sqrt{n_1!\,n_2!}}\,x_{n_1,n_2}=\Phi(z_1)\otimes\varphi^+(z_2),
\label{317}\en
and
\be
\varphi^-(z_1,z_2)=e^{-(|z_1|^2+|z_2|^2)/2}\sum_{n1=0}^\infty\sum_{n2=0}^\infty\frac{z_1^{n_1}\,z_2^{n_2}}{\sqrt{n_1!\,n_2!}}\,x_{n_1,-n_2-1}=\Phi(z_1)\otimes\varphi^-(z_2),
\label{318}\en
together with
\be
\psi^+(z_1,z_2)=e^{-(|z_1|^2+|z_2|^2)/2}\sum_{n1=0}^\infty\sum_{n2=0}^\infty\frac{z_1^{n_1}\,z_2^{n_2}}{\sqrt{n_1!\,n_2!}}\,y_{n_1,n_2}=\Phi(z_1)\otimes\psi^+(z_2),
\label{319}\en
and
\be
\psi^-(z_1,z_2)=e^{-(|z_1|^2+|z_2|^2)/2}\sum_{n1=0}^\infty\sum_{n2=0}^\infty\frac{z_1^{n_1}\,z_2^{n_2}}{\sqrt{n_1!\,n_2!}}\,y_{n_1,-n_2-1}=\Phi(z_1)\otimes\psi^-(z_2).
\label{320}\en
Here we have introduced $\Phi(z_1)=e^{-|z_1|^2/2}\sum_{n=0}^\infty\,\frac{z_1^n}{\sqrt{n!}}\,e_n^{(1)}\in\Hil^{(1)}$ and the other vectors, whose definition is easily derived from formulas (\ref{317})-(\ref{320}) and (\ref{37}), which belong to $\Kc$. The convergence of all these series can be checked, but we will postpone this particular aspect to the most relevant case described in Section \ref{sectadcbcs}. It is clear that, due to (\ref{310}), 
$$
\langle \varphi_\Ac^+(z_1,z_2),\psi_\Ac^-(z_1,z_2)\rangle_{\tilde\Hil}=\langle \varphi_\Ac^-(z_1,z_2),\psi_\Ac^+(z_1,z_2)\rangle_{\tilde\Hil}=0,
$$
while the other scalar products are, in general, non zero. In particular we have,
$$
\langle \varphi_\Ac^\pm(z_1,z_2),\psi_\Ac^\pm(z_1,z_2)\rangle_{\tilde\Hil}=1,
$$
which means that these states are {\em bi-normalized}.  Similarly to (\ref{cs11}) and (\ref{cs12}) we can check the following eigenvalue equations:
\be
\A_{(1)}\varphi^{\pm}(z_1,z_2)=z_1\varphi^{\pm}(z_1,z_2), \qquad \A_{(1)}\psi^{\pm}(z_1,z_2)=z_1\psi^{\pm}(z_1,z_2),
\label{321}\en
\be
\A_\Kc(V)\varphi^+(z_1,z_2)=z_2\varphi^+(z_1,z_2), \qquad \B_\Kc(V)\varphi^-(z_1,z_2)=z_2\varphi^-(z_1,z_2),
\label{322}\en
and

\be
\A_\Kc^\dagger(V)\psi^-(z_1,z_2)=z_2\psi^-(z_1,z_2), \qquad \B_\Kc^\dagger(V)\psi^+(z_1,z_2)=z_2\psi^+(z_1,z_2),
\label{323}\en


\noindent which are in agreement with Figure \ref{fig3} and with the fact that, say, $\A_\Kc(V)$ is a lowering operator for $\tilde\Hil_x^+=\Hil^{(1)}\otimes\Kc_x^+$, while $\B_\Kc(V)$ is a lowering operator for $\tilde\Hil_x^-=\Hil^{(1)}\otimes\Kc_x^-$. Similar conclusions hold for their adjoint. Here we have introduced the sets $\Kc_x^+=\oplus_{n=0}^\infty\Kc_x(n)$, and $\Kc_x^-=\oplus_{n=-\infty}^{-1}\Kc_x(n)$, where $\Kc_x(n)=\overline{l.s.\{x_{m,n}\, m\geq0\}}^{\|.\|}$.

If we now call $\Lc=\Lc_x\cap\Lc_y$, it is easy to check that, $\forall f,g\in\Lc$,
$$
\langle f,g\rangle_{\tilde\Hil}=\frac{1}{\pi^2}\int_{\mathbb{C}}d^2z_1\int_{\mathbb{C}}d^2z_2 \langle f,\varphi^\pm(z_1,z_2)\rangle_{\tilde\Hil}\langle \psi^\pm(z_1,z_2),g\rangle_{\tilde\Hil}=
$$
\be
=\frac{1}{\pi^2}\int_{\mathbb{C}}d^2z_1\int_{\mathbb{C}}d^2z_2 \langle f,\psi^\pm(z_1,z_2)\rangle_{\tilde\Hil}\langle \varphi^\pm(z_1,z_2),g\rangle_{\tilde\Hil}
\label{324}\en
In particular, if $\Lc$ is dense in $\tilde\Hil$ then  $\F_x$ and $\F_y$ are $\Lc$-quasi bases, in the sense of \cite{bagspringer}.

Summarizing, these states work (sufficiently) well: they are eigenstates of our various ladder operators and they resolve the identity under suitable conditions. However, in our opinion they are not really the most convenient vectors to work with in the context of graphene with $V\neq0$. In particular, it is obvious that $\tilde H(V)$ in (\ref{311}) cannot be factorized in terms of the operators $\A_\Kc$, $\B_\Kc$ and their adjoint. For this reason, in the following section, we propose a different definition of bicoherent states which consider this particular aspect of the system.

\subsection{A different class of bicoherent states} \label{sectadcbcs}

The first thing to do is to shift the Hamiltonian $\tilde H(V)$ in such a way zero becomes one of the eigenvalues of this shifted operator. This is important to ensure that the ladder operators we will introduce later behaves as those we have considered all along this paper, which all have some vector which is annihilated by one of them. This would be not so easy, if not impossible, if we try to factorize $\tilde H(V)$ directly, as one can easily check. Going back to (\ref{311}) and (\ref{312}),  recalling that $\tilde H(V)=\1^{(1)}\otimes H(V)$ and using formula (\ref{37}), we have
\be
\left\{
\begin{array}{ll}
	H(V)\,\varphi_{p}=\E_{p}\,\varphi_{p},\\
	H^\dagger(V)\,\psi_{p}=\overline{\E_{p}}\,\psi_{p},
\end{array}
\right.
\label{325}\en
with
$p\in\mathbb{Z}$. We now introduce the following quantities:
\be
\theta_p=\E_{p}-\E_0=\left\{
\begin{array}{ll}
	\frac{2v_F}{\xi}\,(\sqrt{p-V^2}-iV) \hspace{2.6cm}p\geq1,\\
	0, \hspace{5.4cm} p=0,\\
	-\frac{2v_F}{\xi}\,(\sqrt{-p-V^2}+iV),\hspace{1.7cm} p\leq-1.
\end{array}
\right.
\label{326}\en
and the shifted Hamiltonian $h(V)=H(V)-\E_0\1_\Kc$, it follows that $h^\dagger(V)=H^\dagger(V)+\E_0\1_\Kc$ and
\be
\left\{
\begin{array}{ll}
	h(V)\,\varphi_{p}=\theta_{p}\,\varphi_{p},\\
	h^\dagger(V)\,\psi_{p}=\overline{\theta_{p}}\,\psi_{p},
\end{array}
\right.
\label{327}\en
with
$p\in\mathbb{Z}$. As we can see, there is a price to pay: all the eigenvalues of $h(V)$, except one, are complex. However, this is not a major problem for us, since we are working with manifestly non self-adjoint operators from the very beginning, so that reality of eigenvalues is not a constraint.

Let us introduce the following operators:
\be
c_2=\sum_{p=-\infty}^\infty\sqrt{\theta_{p+1}}\,|\varphi_p\rangle \langle \psi_{p+1}|, \qquad d_2=\sum_{p=-\infty}^\infty\sqrt{\theta_{p+1}}\,|\varphi_{p+1}\rangle \langle \psi_{p}|.
\label{328}\en
Here $\sqrt{\theta_p}$ is fixed to be the principal square root\footnote{In other words, given $z=\rho\,e^{i\theta}$, we will always take $\sqrt{z}=\sqrt{\rho}\,e^{i\theta/2}$.} of the complex number $\theta_p$. These operators, and $C_2=\1^{(1)}\otimes c_2$ and $D_2=\1^{(1)}\otimes d_2$ as a consequence, are well defined on {\em large sets}. In fact, in particular, $D(C_2), D(D_2)\supseteq\Lc_x$. It is easy to check now that
\be
h(V)\varphi_{p}=\theta_p\varphi_p=d_2\,c_2\,\varphi_p,
\label{329}\en
$\forall\,p\in\mathbb{Z}$, which means that $h(V)$ can be factorized on $\varphi_p$ in terms of the operators introduced in \eqref{328}. This is because, for all such $p$,
\be
c_2\varphi_p=\sqrt{\theta_p}\varphi_{p-1}, \qquad d_2\varphi_p=\sqrt{\theta_{p+1}}\varphi_{p+1}.
\label{330}\en
The action of $c_2^\dagger$ and $d_2^\dagger$ can be computed using the biorthogonality of $\F_\varphi=\{\varphi_p\}$ and $\F_\psi=\{\psi_p\}$ in $\Kc$, and their completeness in this Hilbert space. We find that
\be
c_2^\dagger\psi_p=\overline{\sqrt{\theta_{p+1}}}\psi_{p+1}, \qquad d_2^\dagger\psi_p=\overline{\sqrt{\theta_{p}}}\psi_{p-1}.
\label{331}\en
As always, given a general sequence of complex numbers $\{\rho_p, \,p\geq0\}$ such that $\rho_0=0$, we define the following quantities: $\rho_0!=1$ and $\rho_n!=\rho_1\rho_2\cdots\rho_n$, $n\geq1$. Hence we can introduce the following vectors:
\be
\eta^+(z)=N(z)\sum_{n=0}^\infty \frac{z^n}{\sqrt{\theta_n!}}\,\varphi_{n}, \qquad \xi^+(z)=N(z)\sum_{n=0}^\infty \frac{z^n}{\overline{\sqrt{\theta_n!}}}\,\psi_{n},
\label{332}\en
as well as
\be
\eta^-(z)=N(z)\sum_{n=0}^\infty \frac{z^n}{\sqrt{\theta_n!}}\,\varphi_{-n-1}, \qquad \xi^-(z)=N(z)\sum_{n=0}^\infty \frac{z^n}{\overline{\sqrt{\theta_n!}}}\,\psi_{-n-1}.
\label{333}\en
Here $N(z)$ ia a suitable normalization, still to be fixed, which as we will see can be taken equal for all these states. What we need to do first is to check the convergence of all these series, which is not granted, a priori. This will be done directly, mimicking the same general strategy proposed in \cite{bagspringer}, Theorem 5.1.1. For instance, focusing on $\eta^+(z)$ and assuming that $N(z)$ is a well defined function (which will be deduced soon), we have
$$
\|\eta^+(z)\|_\Kc\leq|N(z)|\sum_{n=0}^\infty\frac{|z|^n}{|\sqrt{\theta_n!}|}\|\varphi_n\|_\Kc.
$$
First we observe that $|\sqrt{\theta_n!}|=\sqrt{|\theta_n|!}$. Then, to compute $\|\varphi_n\|_\Kc$, we first remember that, see (\ref{33}) and (\ref{38}), $\varphi_0=\begin{pmatrix}
	e_{0}^{(2)} \\
	0
\end{pmatrix}$, while $\varphi_n=\varphi_n^+=K_{n}^{+}(\varphi)\begin{pmatrix}
e_{n}^{(2)} \\
\alpha_{n}^{+}e_{n-1}^{(2)}
\end{pmatrix},$ when $n\geq1$. Then, $\|\varphi_0\|_\Kc=1$, while, for $n\geq1$,
$$
\|\varphi_n\|_\Kc^2=|K_{n}^{+}(\varphi)|^2(1+|\alpha_n^+|^2)=2|K_{n}^{+}(\varphi)|^2,
$$
since from (\ref{34}) one can check that $|\alpha_n^+|^2=1$ for all $n\geq1$. To compute now $|K_{n}^{+}(\varphi)|^2$ we use (\ref{39}). Then we have  $|K_{p}^{\pm}(\varphi)|\,|K_{p}^{\pm}(\psi)|=\sqrt{\frac{p}{4(p-V^2)}}$, which is independent of the choice $\pm$. If we assume, for simplicity and just to fix the ideas, that $|K_{p}^{\pm}(\varphi)|=|K_{p}^{\pm}(\psi)|$, we deduce that
\be
|K_{p}^{\pm}(\varphi)|=|K_{p}^{\pm}(\psi)|=\left(\frac{p}{4(p-V^2)}\right)^{1/4}.
\label{334}\en
Then we conclude that, for all $n\geq1$, 
$$
\|\varphi_n\|_\Kc^2\leq \frac{n}{(n-V^2)}\leq \frac{1}{1-V^2},
$$
as it is easy to check. Now, since $\sqrt{1-V^2}\leq1$, we end up with the following estimate,
\be
\|\varphi_n\|_\Kc^2\leq\frac{1}{(1-V^2)},
\label{335}\en
for all $n\geq0$. This is a {\em good} estimate, since the bound on the right hand side does not depend on $n$. Putting all together we deduce that
 $$
 \|\eta^+(z)\|_\Kc\leq\frac{|N(z)|}{(1-V^2)^{1/2}}\sum_{n=0}^\infty\frac{|z|^n}{|\sqrt{\theta_n!}|},
 $$
which looks essentially like a power series in $|z|$. The radius of convergence of this series can now be found: $\rho=\lim_{n,\infty}\sqrt{|\theta_{n+1}|}=\infty$. This is because, using the expression for $\theta_p$ for $p\geq1$ in (\ref{326}), we find that $|\theta_p|=\frac{2v_F}{\xi}\sqrt{p}$. Similar estimates can be repeated for the other vectors in (\ref{332}) and (\ref{333}), as well as for the states introduced in (\ref{317})-(\ref{320}), which therefore are also well defined for all $z_1,z_2\in\mathbb{C}$. Then we conclude that, if $N(z)$ is a well defined function, $\eta^\pm(z)$ and $\xi^\pm(z)$ are all well defined vectors in $\Kc$, for all $z\in\mathbb{C}$. This implies that their tensor products
\be
\eta^\pm(z_1,z_2)=\Phi(z_1)\otimes\eta^\pm(z_2), \qquad \xi^\pm(z_1,z_2)=\Phi(z_1)\otimes\xi^\pm(z_2),
\label{336}\en
which are clearly the counterparts of (\ref{317})-(\ref{320}), are well defined elements of $\tilde \Hil$ for all $z_1,z_2\in\mathbb{C}$. In particular we have that $\eta^\pm(z_1,z_2)\in\tilde\Hil_x^\pm$, while  $\xi^\pm(z_1,z_2)\in\tilde\Hil_y^\pm$.

In order to fix the function $N(z)$ we next require that $\langle\eta^\pm(z_1,z_2),\xi^\pm(z_1,z_2)\rangle_{\tilde\Hil}=1$. Since $\langle\Phi(z_1),\Phi(z_1)\rangle_{(1)}=1$, this implies that
\be
|N(z_2)|^2\sum_{n=0}^\infty\frac{|z_2|^{2n}}{|\theta_n|!}=1 \qquad \Rightarrow \qquad N(z_2)=\left(\sum_{n=0}^\infty\frac{|z_2|^{2n}}{|\theta_n|!}\right)^{-1/2},
\label{337}\en
with a proper choice of the phase for $N(z_2)$. The series converge for all $z_2\in\mathbb{C}$. This is because its radius of convergence
is infinite. This can be checked as we did above. Of course, formula (\ref{337}) implies that $N(z_2)$ only depends on $|z_2|$. For this reason, we write $N(|z_2|)$ from now on. Needless to say $\sum_{n=0}^\infty\frac{|z_2|^{2n}}{|\theta_n|!}\neq0$ for all $z_2\in\mathbb{C}$, since the series is always strictly larger than 1.

As we have already noticed for the states in (\ref{317})-(\ref{320}), the {\em plus} states are orthogonal to the {\em minus} ones. In other words, we have that $\langle\eta^\pm(z_1,z_2),\xi^\mp(z_1,z_2)\rangle_{\tilde\Hil}=0$.

Going on with our analysis on the states in (\ref{336}) we observe that the following ladder equations hold:
\be
\left\{
\begin{array}{ll}
	C_2\,\eta^+\,(z_1,z_2)=z_2\,\eta^+\,(z_1,z_2), \qquad C_2^\dagger\,\xi^-\,(z_1,z_2)=z_2\,\xi^-\,(z_1,z_2),\\
	D_2\,\eta^-\,(z_1,z_2)=z_2\,\eta^-\,(z_1,z_2), \qquad D_2^\dagger\,\xi^+\,(z_1,z_2)=z_2\,\xi^+\,(z_1,z_2),
\end{array}
\right.
\label{338}\en
which are in agreement with our decomposition of $\tilde\Hil$, and with our interpretation of the various ladder operators. We have further that
\be
\A_{(1)}\eta^\pm\,(z_1,z_2)=z_1\,\eta^\pm\,(z_1,z_2), \qquad \A_{(1)}\xi^\pm\,(z_1,z_2)=z_1\,\xi^\pm\,(z_1,z_2).
\label{339}\en
which are the counterpart, here, of the analogous equations in (\ref{321}).

To conclude our analysis of the states in (\ref{336}) we write $z_j=r_j\,e^{i\theta_j}$, $j=1,2$, and we introduce the (standard) measure $d\nu(z_1,\overline z_1)=\frac{1}{\pi}r_1\,dr_1\,d\theta_1$ and a second measure $d\nu(z_2,\overline z_2)=d\lambda_2(r_2)\,d\theta_2$, where $r_j>0$ and $\theta_j\in[0,2\pi[$, $j=1,2$, and $d\lambda_2(r_2)$ is assumed to satisfy the following moment problem
\be
\int_0^\infty
d\lambda_2(r_2) r_2^{2n} N^2(r_2)=\frac{|\theta_n|!}{2\pi},
\label{340}\en
$\forall n\geq0$.
We should stress that this is really an assumption here, since we have no explicit form for $d\lambda_2(r_2)$ so far. Constructing a similar measure is work in progress. However, if such a measure exists, then we can write
\be
\int_{\mathbb{C}}d\nu(z_1,\overline z_1)\int_{\mathbb{C}}d\nu(z_2,\overline z_2)\langle \tilde f,\eta^+(z_1,z_2)\rangle_{\tilde\Hil}\langle \xi^+(z_1,z_2), \tilde g\rangle_{\tilde\Hil}=\sum_{n_=0}^\infty\sum_{p=0}^\infty\langle \tilde f,x_{n,p}\rangle_{\tilde\Hil}\langle y_{n,p}, \tilde g\rangle_{\tilde\Hil}
\label{341}\en
and
\be
\int_{\mathbb{C}}d\nu(z_1,\overline z_1)\int_{\mathbb{C}}d\nu(z_2,\overline z_2)\langle \tilde f,\xi^+(z_1,z_2)\rangle_{\tilde\Hil}\langle \eta^+(z_1,z_2), \tilde g\rangle_{\tilde\Hil}
=\sum_{n_=0}^\infty\sum_{p=0}^\infty\langle \tilde f,y_{n,p}\rangle_{\tilde\Hil}\langle x_{n,p}, \tilde g\rangle_{\tilde\Hil}.
\label{342}\en
Of course the right hand sides of these formulas both return $\langle \tilde f, \tilde g\rangle_{\tilde\Hil}$ if $\F_x$ and $\F_y$ are $\tilde\D$-quasi bases, for some $\tilde\D\subseteq\tilde\Hil$, and $\tilde f,\tilde g\in\tilde\D$. Similar formulas can be found for the pair $(\eta^-(z_1,z_2),\xi^-(z_1,z_2))$.

\section{Chemical potential: $V>1$}\label{sect4b}
In what follows we extend our analysis to the case $V>1$, by considering first the case in which the second index in $x_{n,p}$ and $y_{n,p}$ is such that $|p|\neq V^2$ for all $p\in\mathbb{Z}$. We will comment on the case when $|p|= V^2$ for some $p$ at the end of the section.

The main difference now relies in the fact that the eigenvalues of $\tilde{H}(V)$ and $\tilde{H}^\dagger(V)$ defined in \eqref{312} are complex when $|p|<V^2$  extending, as shown in \cite{BagaHatano_2016}, the width of the region where the  $\mathcal{PT}$ symmetry is broken. This has an important consequence on the biorthonormality conditions for the eigenvectors of our Hamiltonians, when compared to the case $V<1$.
In particular,  for any $q\geq0$, we have the following biorthonormal conditions that depends whether we are in the $\mathcal{PT}$-broken or unbroken  region.\\
\noindent
For \( 1\leq p < V^2 \) (\textit{broken} region), we obtain:
\be
\langle \varphi_p^{\pm}, \psi_p^{\pm} \rangle = 0,   \qquad
\langle  \varphi_p^{\pm}, \psi_q^{\mp} \rangle  = \delta_{p,q},
\label{41}\en
while, for $p > V^2$ (\textit{unbroken} region) 
\be 
\langle  \varphi_p^{\pm}, \psi_q^{\mp}  \rangle  = 0 \qquad
\langle  \varphi_p^{\pm}, \psi_q^{\pm}  \rangle  = \delta_{p,q}.
\en 
with normalization factors given by  \be
\overline{K_{p}^{\pm}(\varphi)}\,K_{p}^{\pm}(\psi)=\frac{p}{2\left(p-V^2\pm i V\sqrt{p-V^2}\right)}.
\en
This implies that extending the construction of the ladder operators proposed for the case $V<1$, requires a careful balance between the $+$ and $-$ vectors, depending on whether we are in the broken or unbroken phase. Considering this, it might be more advantageous to rearrange the sets in a suitable manner. One such a possibility is the following: 
\be\label{44}
\tilde{\psi}_{p}^{\pm}= 
\begin{cases}
	\psi_p^{\mp} \qquad \text{for} \qquad p<V^2, \\
	\psi_p^{\pm} \qquad \text{for} \qquad p>V^2.
\end{cases}
\en
and, consequently,
\be
x_{n,p}=e_n^{(1)}\otimes\varphi_p, \qquad \tilde{y}_{n,p}=e_n^{(1)}\otimes\tilde{\psi}_p,
\label{45}\en
where $\varphi_p$ and $\tilde{\psi}_p$ are specified as in \eqref{38}, just replacing ${\psi}^{\pm}_p$ with $\tilde{\psi}^{\pm}_p$.
Now the two sets $\F_x=\{x_{n,p}\}$ and $\F_{\tilde{y}}=\{\tilde{y}_{n,q}\}$ are biorthonormal in $\tilde\Hil$, as they satisfy
\be
\langle x_{n,p},\tilde{y}_{m,q}\rangle_{\tilde\Hil}= \langle e_n^{(1)},e_m^{(1)}\rangle_{1} \langle\varphi_p,\tilde\psi_q\rangle_{\Kc}=\delta_{n,m}\delta_{p,q}.
\label{46}\en
%
%

We now discuss which are the differences with what we have found before, focusing only on the more interesting case, i.e. on the bicoherent states introduced in Section \ref{sectadcbcs}.

The bicoherent states extend those defined in \eqref{332}-\eqref{333} and in \eqref{336}:
\be
\xi^+(z_1,z_2)=\Phi(z_1)\otimes \left(N^+(z_2)\sum_{n=0}^\infty \frac{z_2^n}{\overline{\sqrt{\theta_n!}}}\,\tilde\psi_{n}\right),\quad \xi^-(z_1,z_2)=\Phi(z_1)\otimes \left(N^-(z_2)\sum_{n=0}^\infty \frac{z_2^n}{\overline{\sqrt{\theta_n!}}}\,\tilde\psi_{-n-1}\right),
\label{410}\en
where $N^\pm(z_2)$ are as usual chosen to satisfy $\langle\eta^\pm(z_1,z_2),\xi^\pm(z_1,z_2)\rangle_{\tilde\Hil}=1$. As in the case $V<1$ one need to check first the convergence of the series
$
\sum_{n=0}^\infty \frac{z_2^n}{\overline{\sqrt{\theta_n!}}}\,\tilde\psi_{n}
$ and $
\sum_{n=0}^\infty \frac{z_2^n}{\overline{\sqrt{\theta_n!}}}\,\tilde\psi_{-n-1}
$. We will show that these series indeed converge in all the complex plane. For that, we now rewrite the first series as follows (a similar strategy can be adopted for the second series):
\be
N^+(z_2)\left[\left(\sum_{n=0}^{[V^2]} \frac{z_2^n}{\overline{\sqrt{\theta_n!}}}\,\tilde\psi_{n}\right)+
\left(\sum_{n=[V^2]+1}^{\infty} \frac{z_2^n}{\overline{\sqrt{\theta_n!}}}\,\tilde\psi_{n}\right)\right]
\en
where $[V^2]$ denotes the integer part of $V^2$. In this decomposition, we see that the finite sum comes from the \textit{broken phase}, while the infinite series comes from the \textit{unbroken phase}. Convergence of the first (finite) sum is obvious. The convergence of the infinite series can be deduced slightly modifying what we have done for $V<1$. In fact,  for all $n\geq[V^2]+1$, we can check that
$$
\|\tilde\psi_n\|_\Kc^2\leq \frac{n}{n-V^2}\leq \frac{[V^2]+1}{[V^2]+1-V^2},
$$
 which, as before, allows to prove the convergence of the series for any $z_2\in\mathbb{C}$.

We conclude this section discussing what happens in the case $n=V^2$ for some $n$. As shown in \cite{BagaHatano_2016}  the completeness properties of the sets  $\F_x=\{x_{n,p}\},\F_{{y}}=\{\tilde{y}_{n,p}\}$ fails. This is a consequence of the fact that, for  $n=V^2$, we have $\alpha^+_{n}=\alpha^-_{n}=-1$ which implies, in turns, that $\varphi^+_n=\varphi^-_n$ and $\tilde\psi^+_n=\tilde\psi^-_n$, which also have {coincident eigenvalue 0}. Hence  $n=V^2$ defines an \textit{exceptional point}.
Moveover
 \be \langle\varphi_n,\tilde\psi_n\rangle_{\Kc}=0,\en
so that the ladder operators $C_2$ and $D_2$, and their adjoints, can no longer be constructed as we have done before, see (\ref{328}) and what follows. Of course, we have the same difficulties for the operators in \eqref{313} and \eqref{315}. This implies that our definitions of bicoherent states do not work anymore. This is the typical critical situation that appears in presence of exceptional points, so that it is not a big surprise we meet these problems also here.


\subsection{Gain and loss phenomena for large $V$}
In this section we show some plots which highlights a critical behaviour when the region of $\mathcal{P}\mathcal{T}$ broken symmetry increases.
Indeed, when $V$ is very large, virtually $V\rightarrow\infty$,   the region where the $\mathcal{P}\mathcal{T}$-symmetry is broken enlarges, and we observe a critical behaviours for the coefficients $\alpha^{\pm}_{n}$, that is
\be\label{vinfunbro}
|\alpha^{+}_{n}|\rightarrow0,\quad |\alpha^{-}_{n}|\rightarrow+\infty,
\en 
and more in general, when  $V^2\gg p$ we have the approximations 
\bea\label{2p24}
\varphi_{n,p}^{+},\psi_{n,p}^{-}\propto  \begin{pmatrix}
	e_{n,p} \\
	0
\end{pmatrix},\quad \varphi_{n,p}^{-}, \psi_{n,p}^{+}\propto  \begin{pmatrix}
	0 \\
	e_{n,p-1}
\end{pmatrix}.
\ena
Hence the presence of a large $\mathcal{PT}$-\textit{broken} region is characterized by the concentration of probability densities predominantly in either the first or second component, depending on the specific sets under consideration.  As extensively documented in the scientific literature,  \cite{benbook}, this phenomenon is attributed to the typical gain and loss dynamics inherent to the $\mathcal{PT}$-\textit{broken} symmetry. {In the context of the two-dimensional vector representing the graphene layer state, this can be explained  by a mechanisms that produces a gain (amplification) of the  electron density in one sublattice of the layer, and the loss (attenuation) in the other sublattice.}
Such an instability is intrinsically linked to the transition towards non-real eigenvalues.
Conversely, in the $\mathcal{PT}$-\textit{unbroken} region, the system exhibits a balanced behaviour characterized by a constant flow of electron and an equilibrium between the two sublattices. This balanced state is linked to the reality of the eigenvalues.

\begin{figure*}[!tbp]\begin{center}
		\subfigure[$|\Phi^+_\mathcal{A}(z_1,z_2)|^2$]{\includegraphics[width=.32\textwidth]{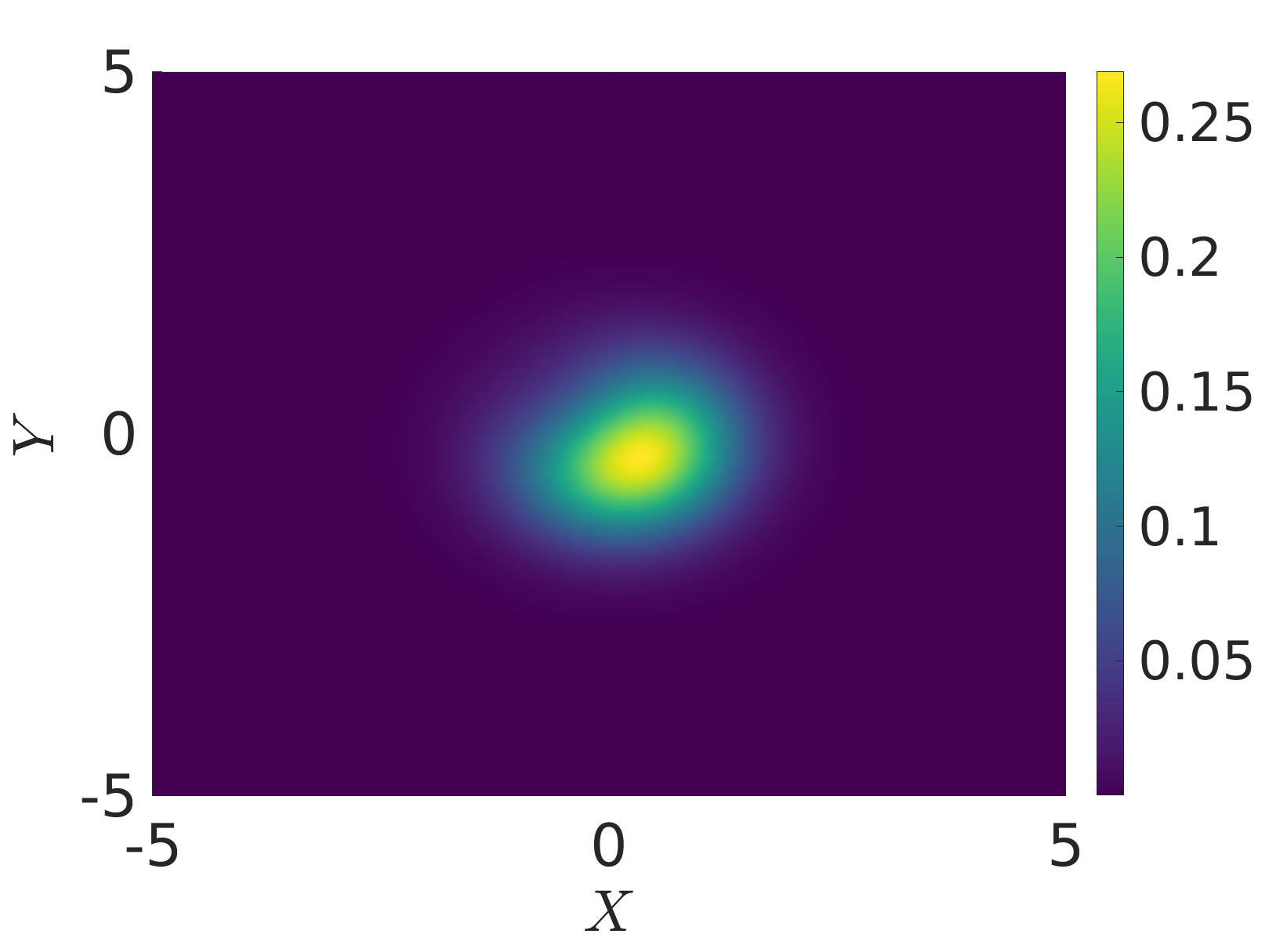}\label{fig::taupf}}
		\hspace*{-0cm}\subfigure[$|\Phi^+_\mathcal{A}(z_1,z_2)\{1\}|^2$]{\includegraphics[width=.32\textwidth]{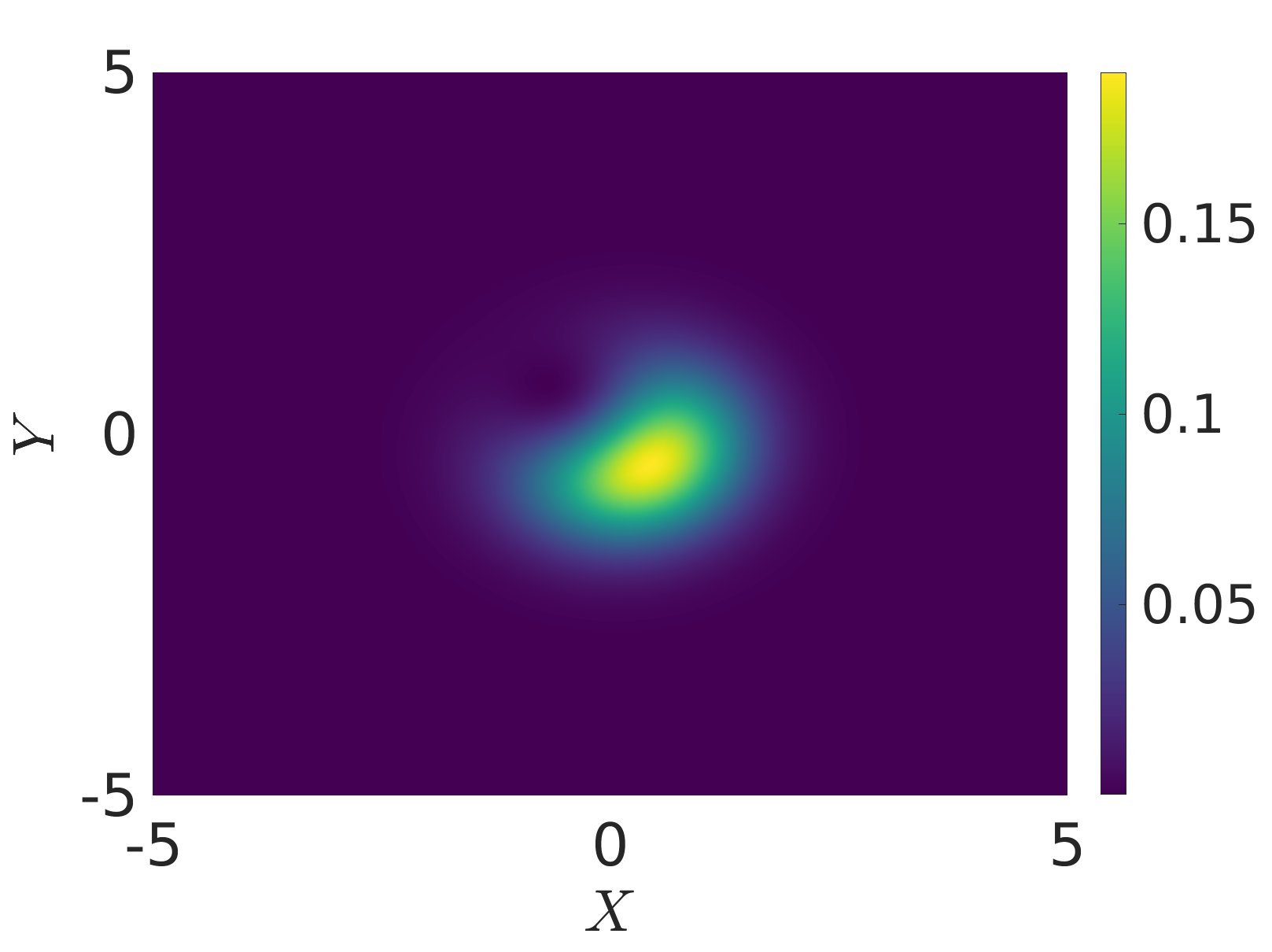}\label{fig::taupf1}}
			\hspace*{-0cm}\subfigure[$|\Phi^+_\mathcal{A}(z_1,z_2)\{2\}|^2$]{\includegraphics[width=.32\textwidth]{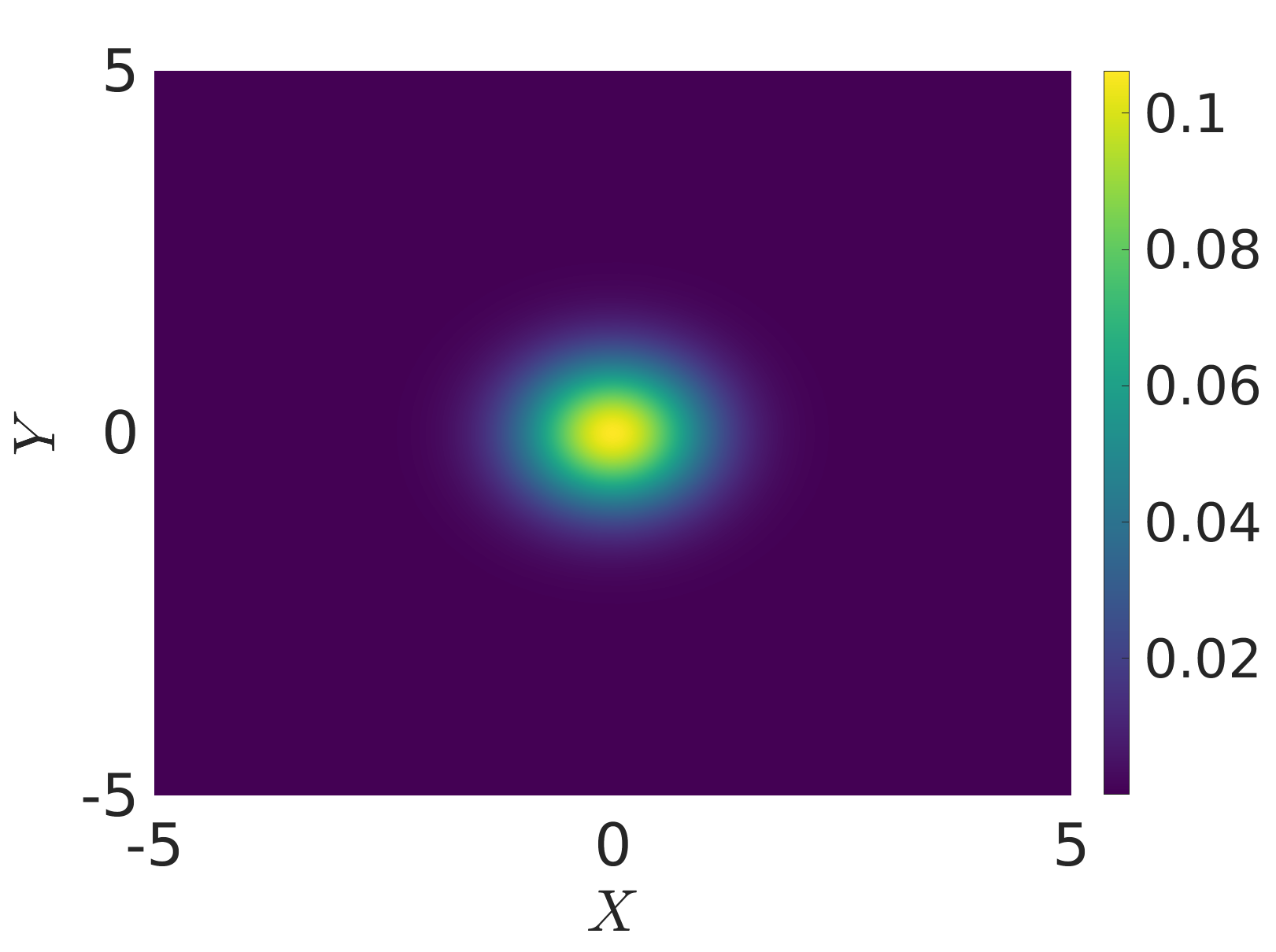}\label{fig::taupf2}}
		\caption{Probability density for $|\Phi^+_\mathcal{A}(z_1,z_2)|^2$ (a) for its first component $|\Phi^+_\mathcal{A}(z_1,z_2)\{1\}|^2$ (b) and its second component $|\Phi^+_\mathcal{A}(z_1,z_2)\{2\}|^2$ (c) . Parameters are $V=0$ and $z_1=0,z_2=1-i$.}
		\label{fig33}
	\end{center}
\end{figure*}

\begin{figure*}[!tbp]\begin{center}
		\subfigure[$|\varphi^+(z_1,z_2)|^2$]{\includegraphics[width=.32\textwidth]{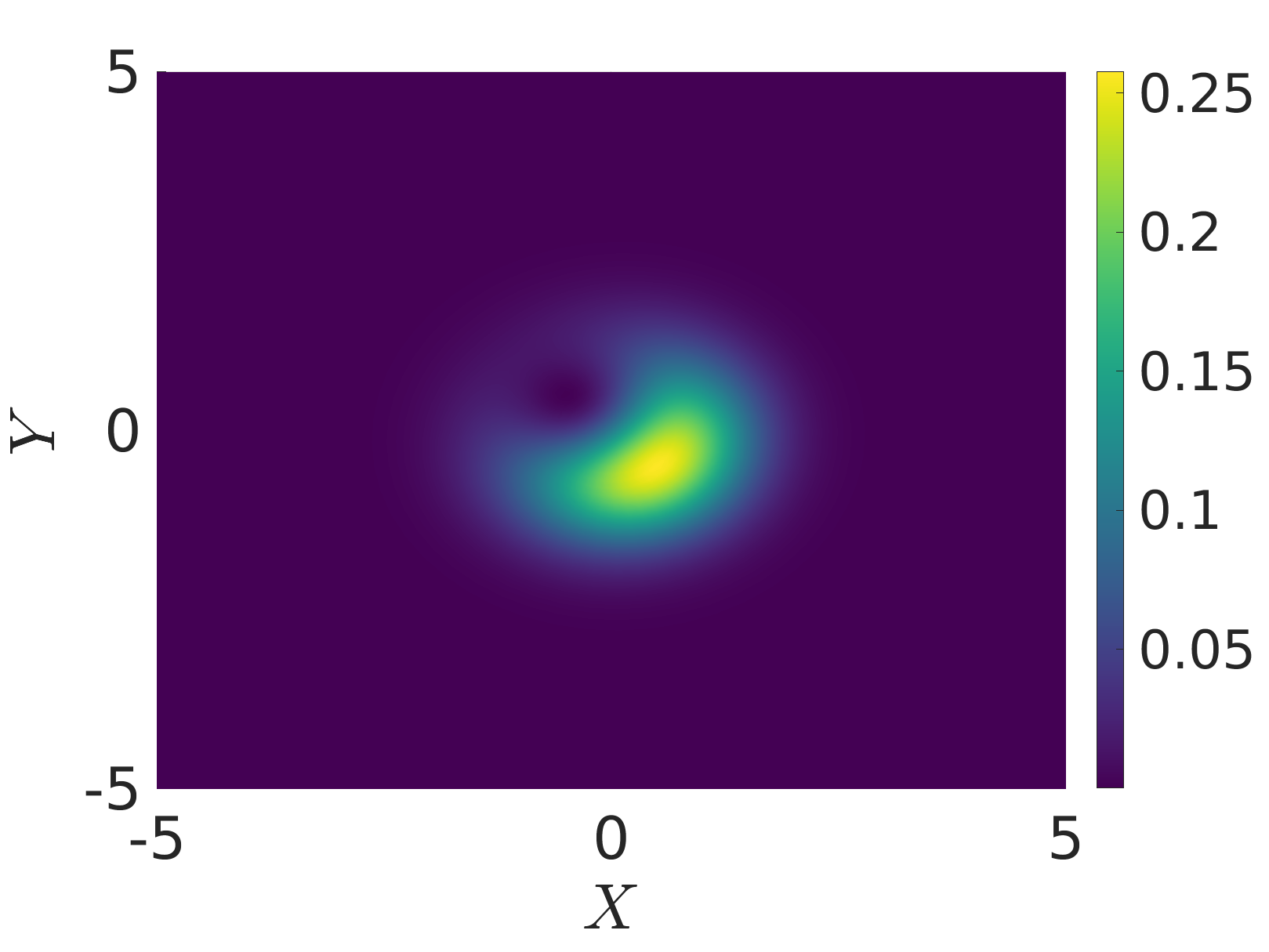}\label{fig::phipf}}
		\hspace*{-0cm}\subfigure[$|\varphi^+(z_1,z_2)\{1\}|^2$]{\includegraphics[width=.32\textwidth]{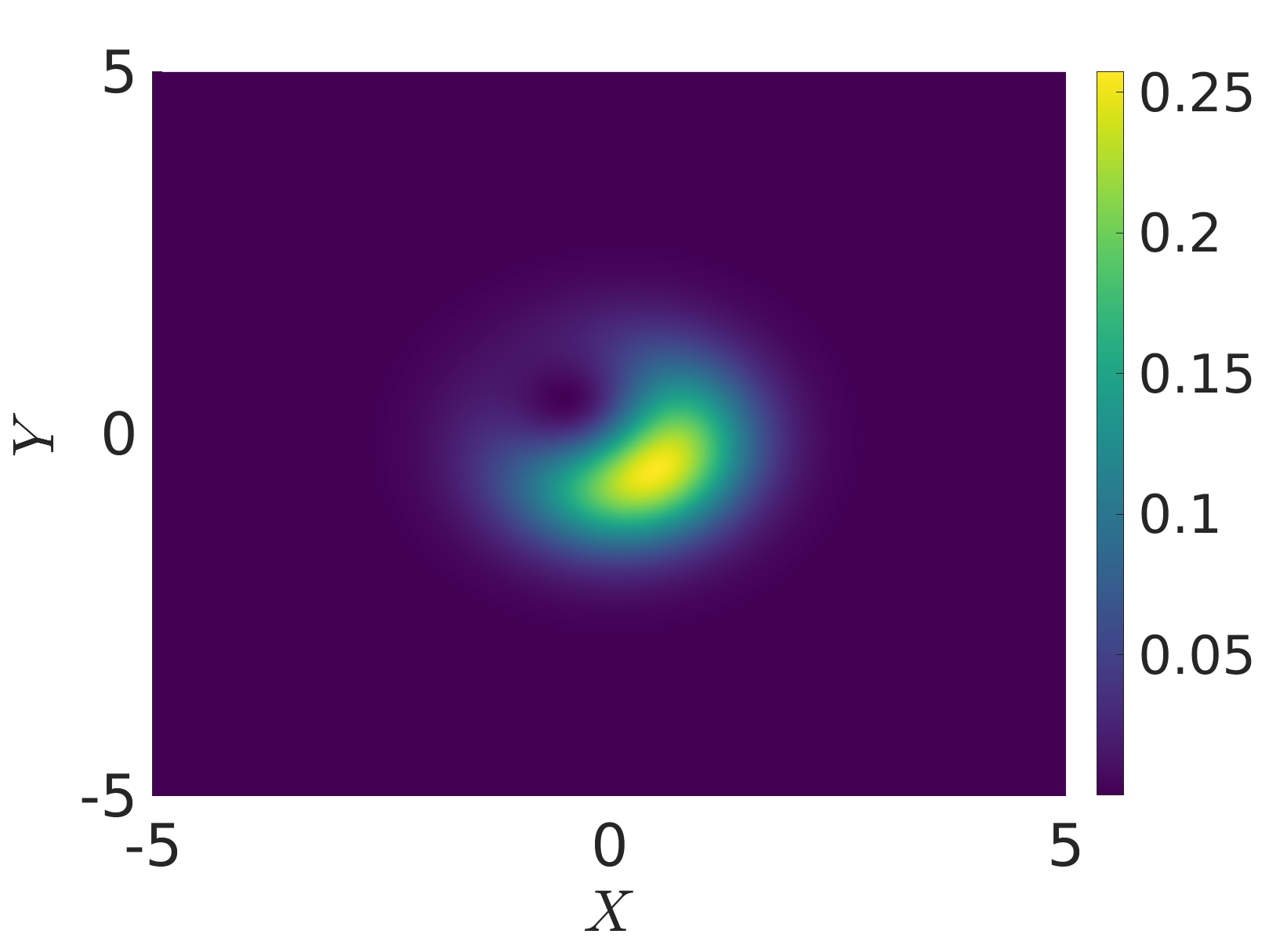}\label{fig::phipf1}}
		\hspace*{-0cm}\subfigure[$|\varphi^+(z_1,z_2)\{2\}|^2$]{\includegraphics[width=.32\textwidth]{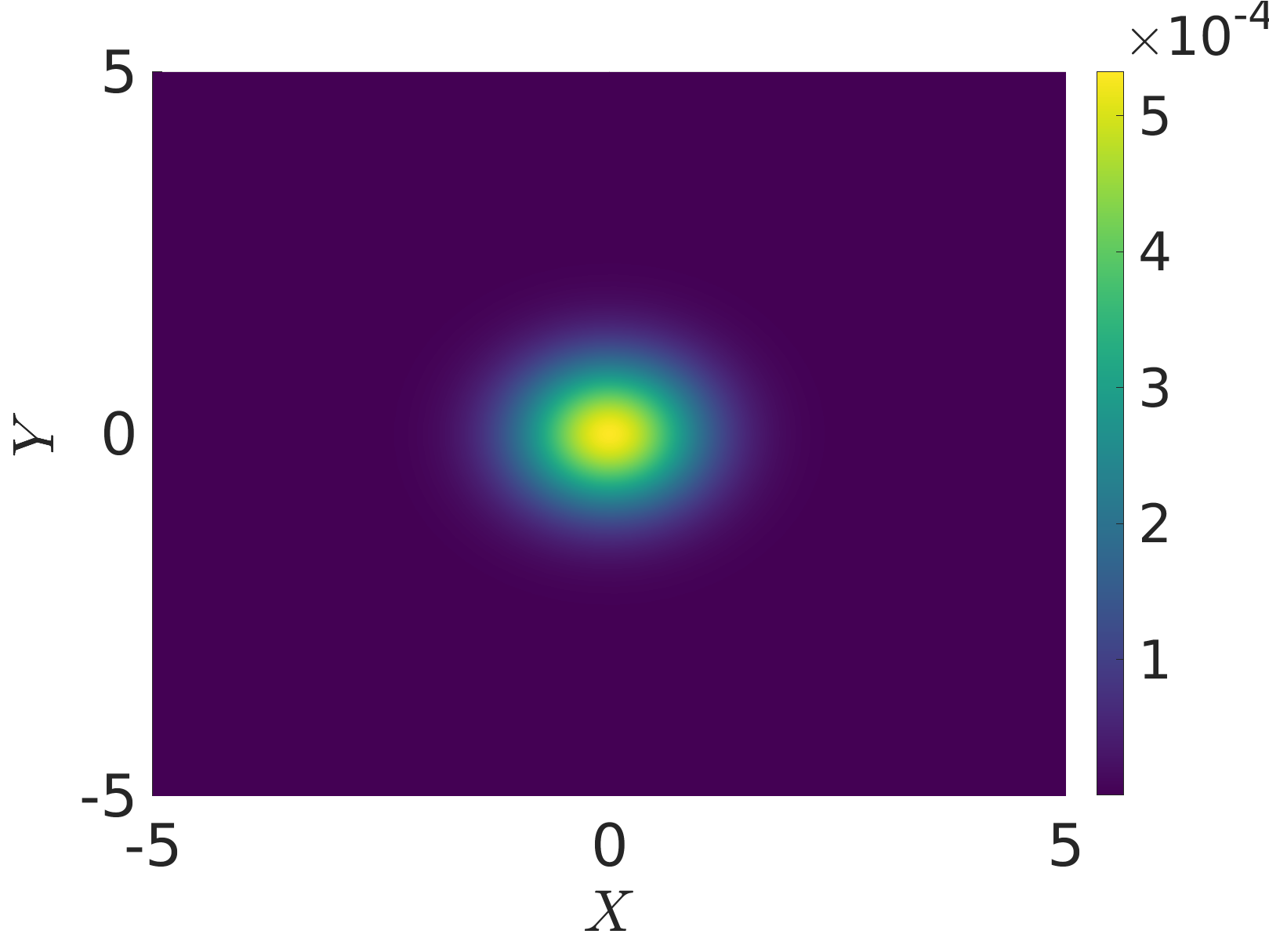}\label{fig::phipf2}}
		\caption{Probability density for $|\varphi^+(z_1,z_2)|^2$ (a) for its first component $|\varphi^+(z_1,z_2)\{1\}|^2$ (b) and its second component $|\varphi^+(z_1,z_2)\{2\}|^2$ (c) . Parameters are $V=9.5$ and $z_1=0,z_2=1-i$.}
		\label{fig4}
	\end{center}
\end{figure*}

\begin{figure*}[!tbp]\begin{center}
		\subfigure[$|\psi^-(z_1,z_2)|^2$]{\includegraphics[width=.32\textwidth]{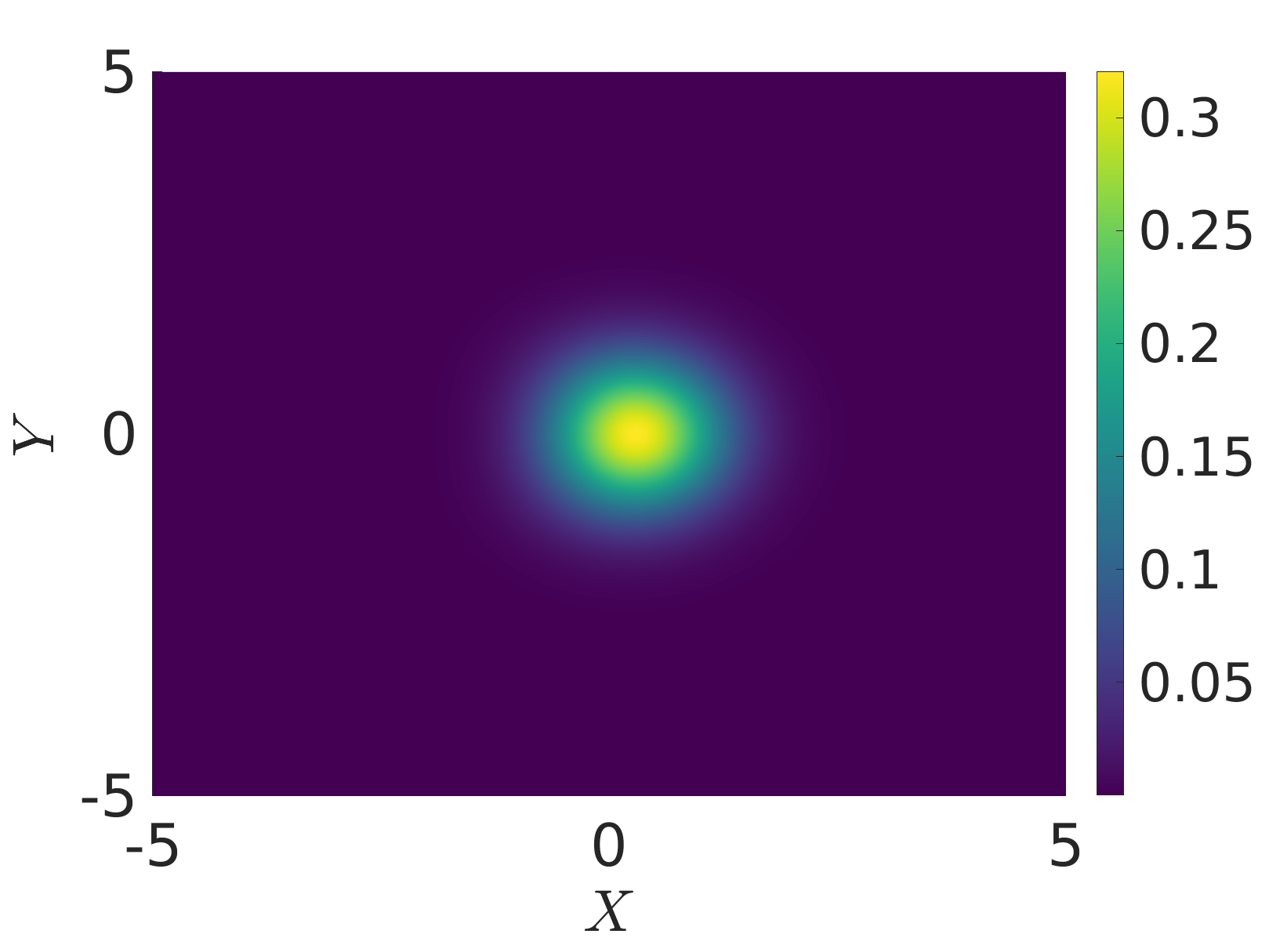}\label{fig::psimf}}
		\hspace*{-0cm}\subfigure[$|\psi^-(z_1,z_2)\{1\}|^2$]{\includegraphics[width=.32\textwidth]{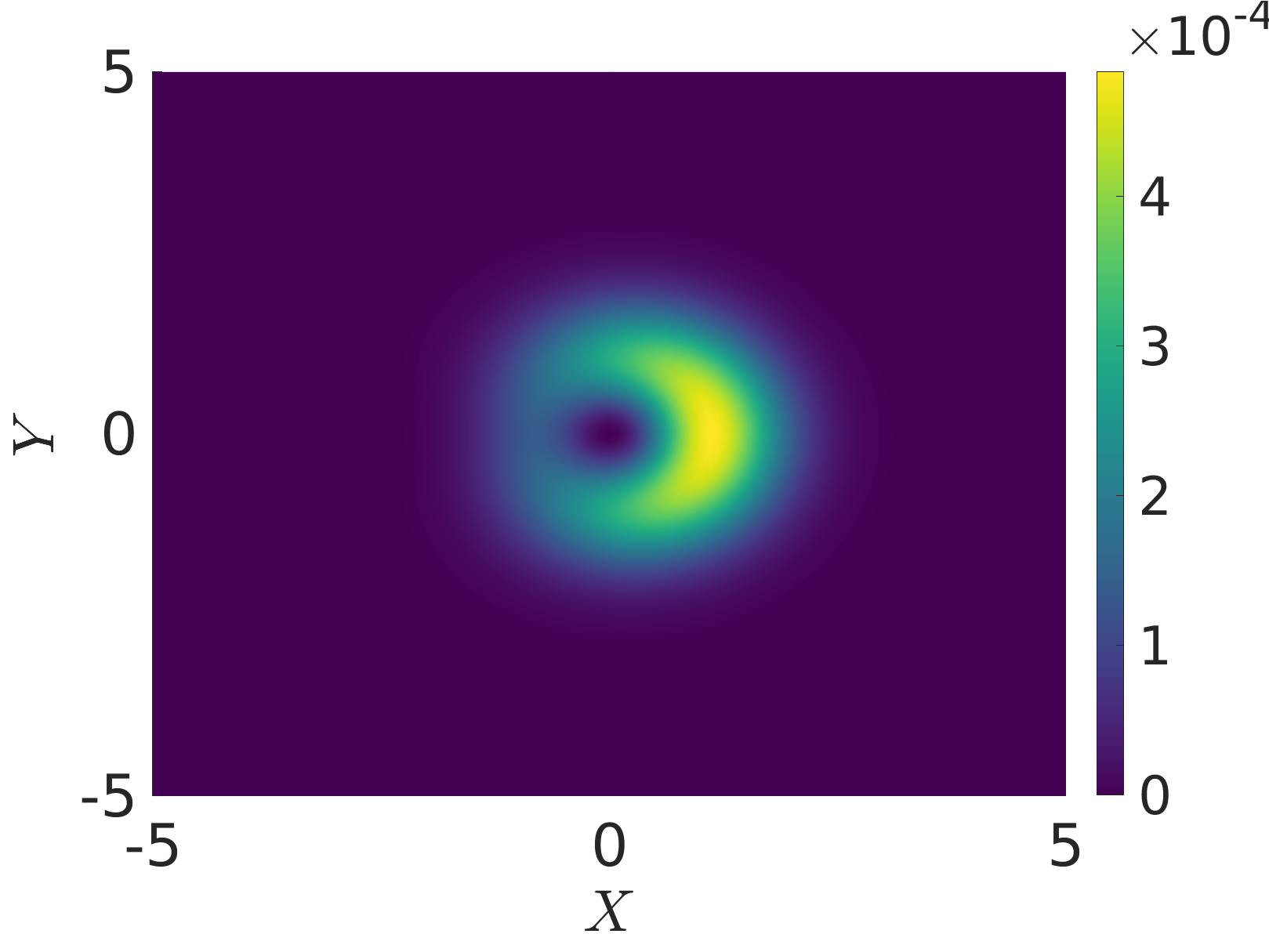}\label{fig::psimf1}}
		\hspace*{-0cm}\subfigure[$|\psi^-(z_1,z_2)\{2\}|^2$]{\includegraphics[width=.32\textwidth]{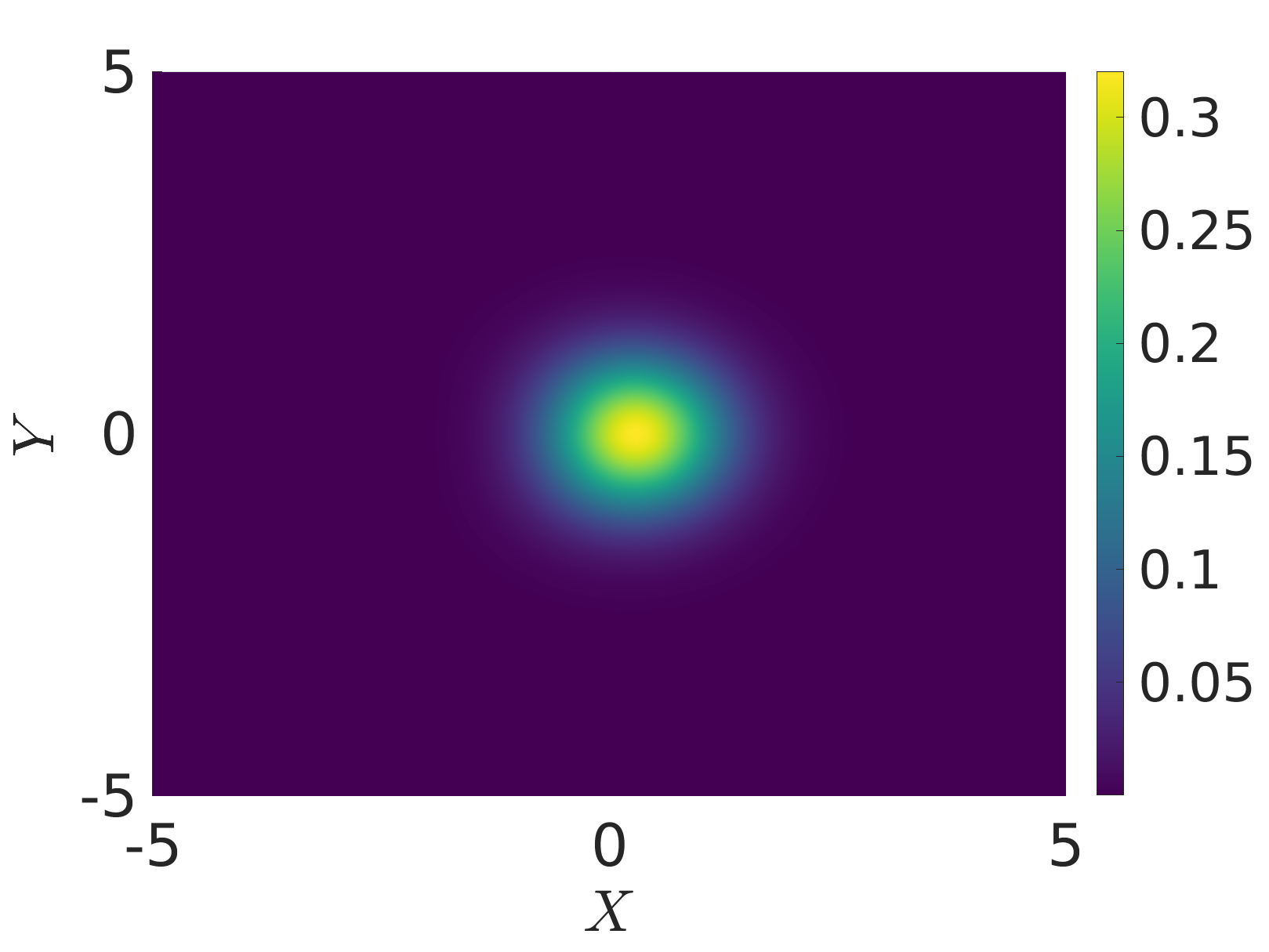}\label{fig::psimf2}}
		\caption{Probability density for $|\psi^-(z_1,z_2)|^2$ (a) for its first component $|\psi^-(z_1,z_2)\{1\}|^2$ (b) and its second component $|\psi^-  (z_1,z_2)\{2\}|^2$ (c) . Parameters are $V=9.5$ and $z_1=0,z_2=1-i$.}
		\label{fig5}
	\end{center}
\end{figure*}

\begin{figure*}[!tbp]\begin{center}
		\subfigure[$|\eta^+(z_1,z_2)|^2$]{\includegraphics[width=.32\textwidth]{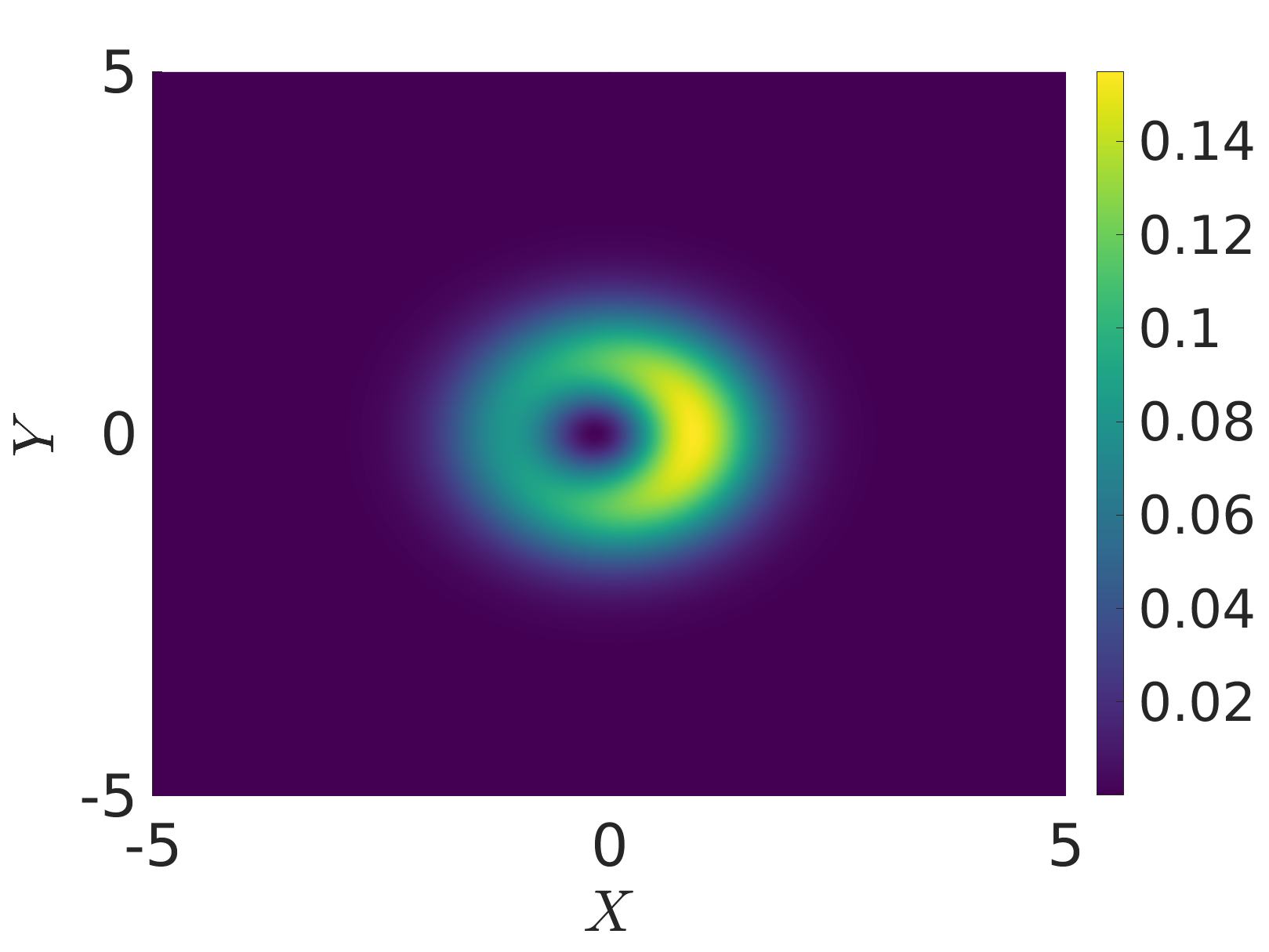}\label{fig::phipf}}
		\hspace*{-0cm}\subfigure[$|\eta^+(z_1,z_2)\{1\}|^2$]{\includegraphics[width=.32\textwidth]{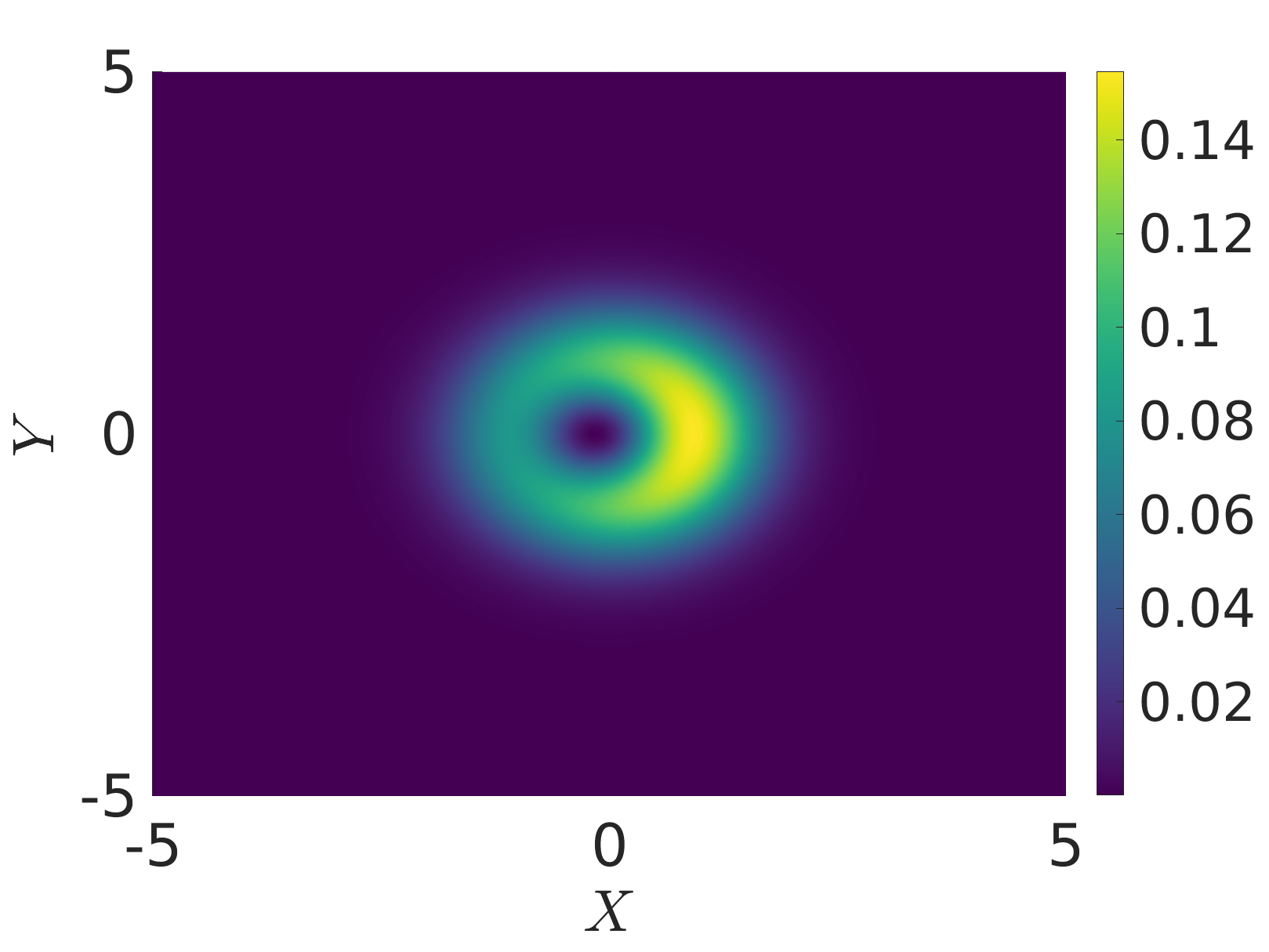}\label{fig::phipf1}}
		\hspace*{-0cm}\subfigure[$|\eta^+(z_1,z_2)\{2\}|^2$]{\includegraphics[width=.32\textwidth]{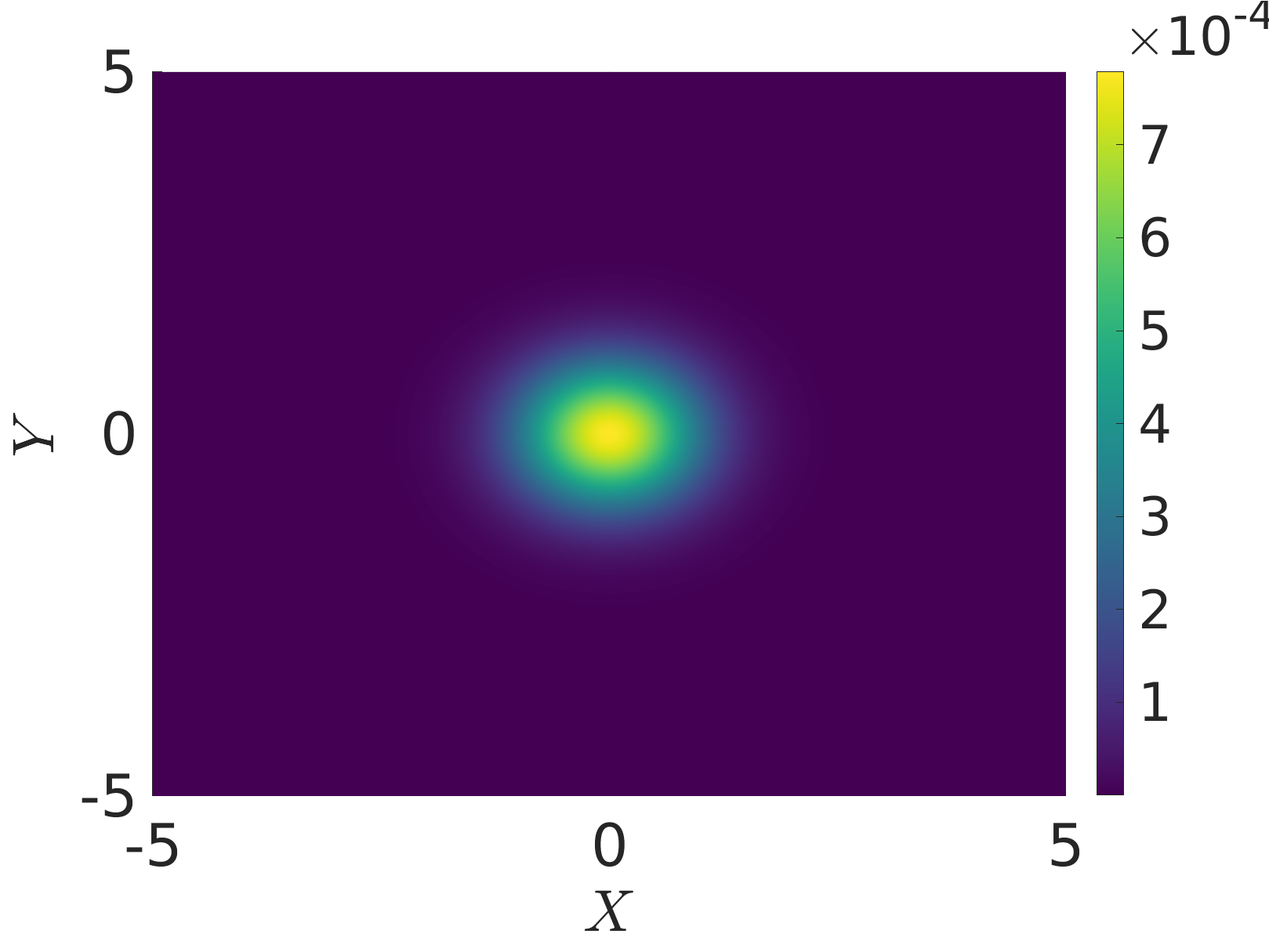}\label{fig::phipf2}}
		\caption{Probability density for $|\eta^+(z_1,z_2)|^2$ (a) for its first component $|\eta^+(z_1,z_2)\{1\}|^2$ (b) and its second component $|\eta^+(z_1,z_2)\{2\}|^2$ (c) . Parameters are $V=9.5$ and $z_1=0,z_2=1-i$.}
		\label{fig7}
	\end{center}
\end{figure*}

\begin{figure*}[!tbp]\begin{center}
		\subfigure[$|\xi^-(z_1,z_2)|^2$]{\includegraphics[width=.32\textwidth]{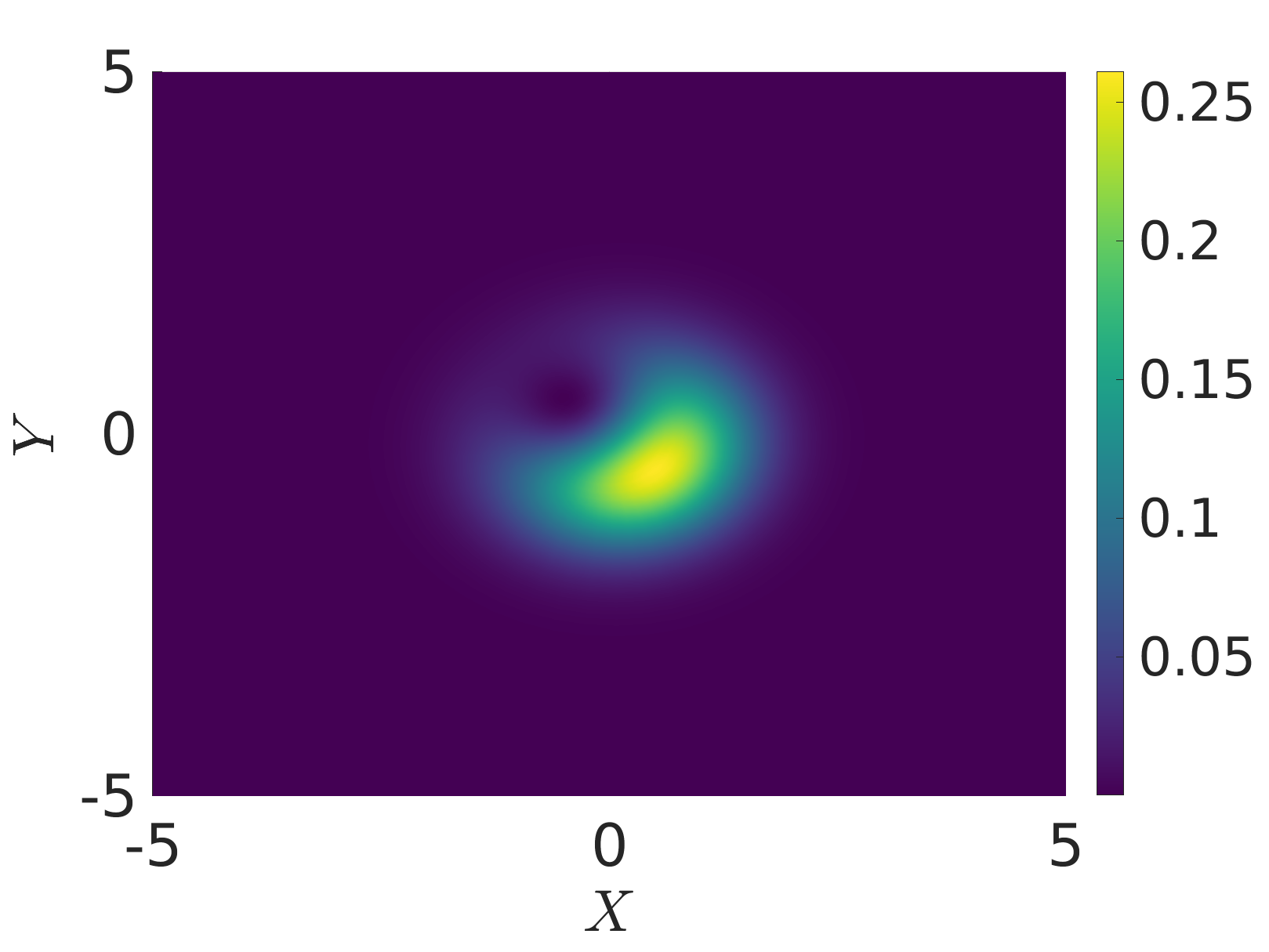}\label{fig::psimf}}
		\hspace*{-0cm}\subfigure[$|\xi^-(z_1,z_2)\{1\}|^2$]{\includegraphics[width=.32\textwidth]{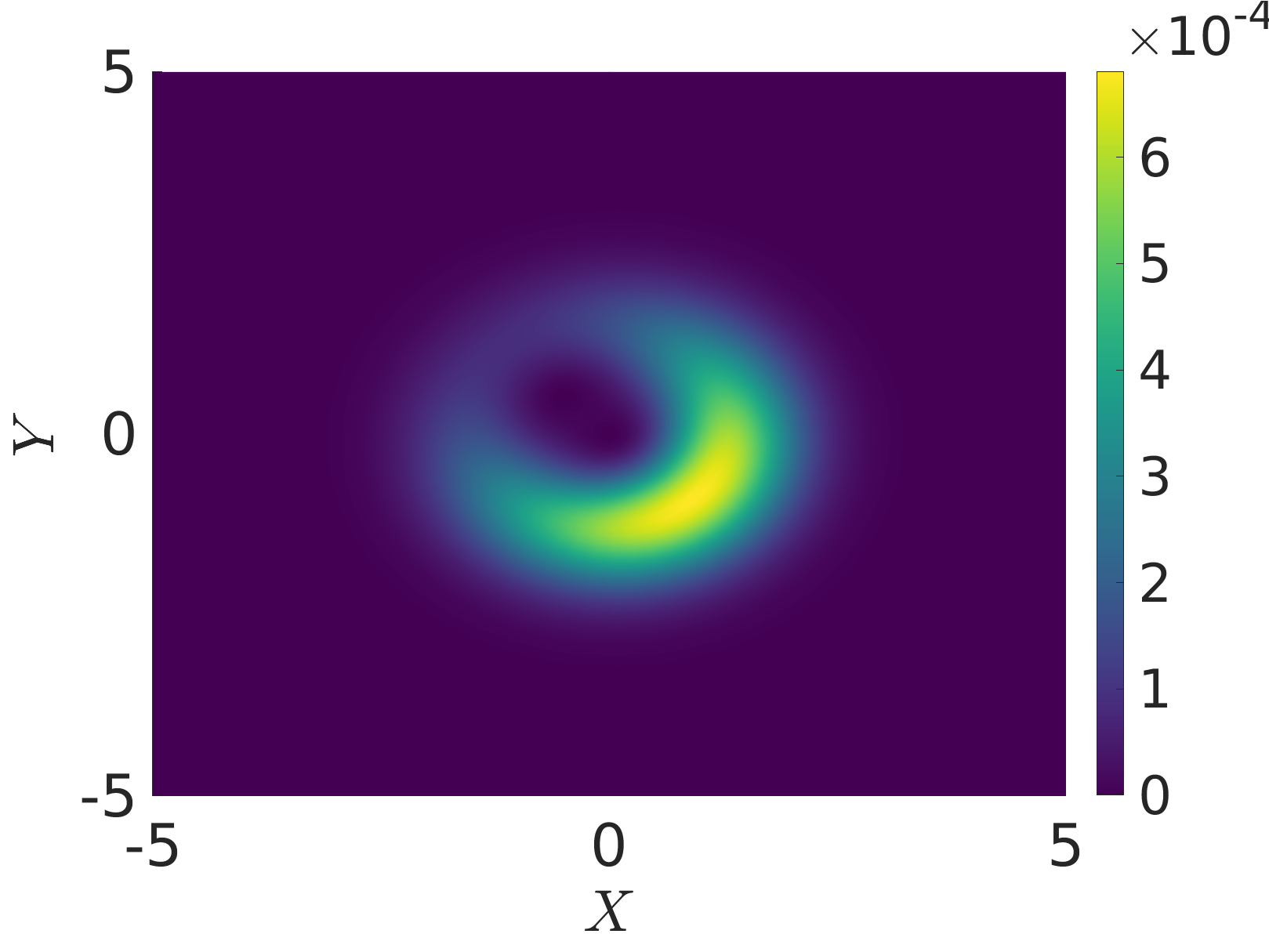}\label{fig::psimf1}}
		\hspace*{-0cm}\subfigure[$|\xi^-(z_1,z_2)\{2\}|^2$]{\includegraphics[width=.32\textwidth]{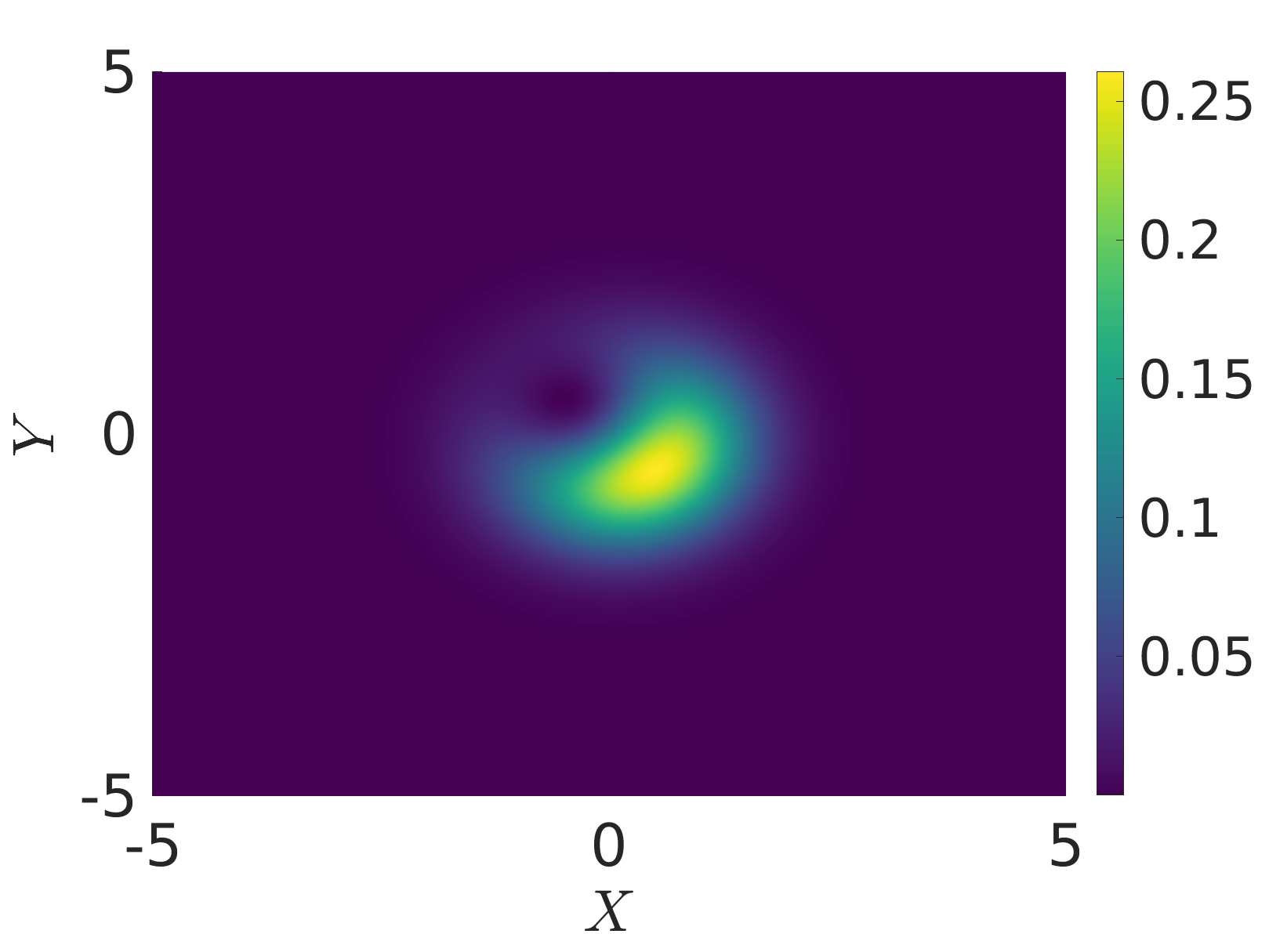}\label{fig::psimf2}}
		\caption{Probability density for $|\xi^-(z_1,z_2)|^2$ (a) for its first component $|\xi^-(z_1,z_2)\{1\}|^2$ (b) and its second component $|\xi^-  (z_1,z_2)\{2\}|^2$ (c) . Parameters are $V=9.5$ and $z_1=0,z_2=1-i$.}
		\label{fig8}
	\end{center}
\end{figure*}

To highlight this critical behavior, we show several plots that support this explanation for the case $V=9.5$. In Figure \ref{fig33}, we illustrate the probability densities for the standard coherent state $|\Phi^+_\mathcal{A}(z_1,z_2)|^2$ at $V=0$ with $z_1=0$ and $z_2=1-i$. We also depict the probability densities $|\Phi^+_\mathcal{A}(z_1,z_2)\{1\}|^2$ and $|\Phi^+_\mathcal{A}(z_1,z_2)\{2\}|^2$ for its first and second components\footnote{Given a vector $v=(v_1,v_2)$ we simply put $v\{1\}=v_1,\, v\{2\}=v_2$.}, respectively. Figures \ref{fig4} and \ref{fig5} contain analogous plots for the bicoherent states $\varphi^+(z_1,z_2)$ and $\psi^-(z_1,z_2)$. It is more appropriate to compare the coherent states $\varphi^+(z_1,z_2)$ and $\psi^-(z_1,z_2)$ because of the order of the vectors $\tilde\psi^{\pm}_n$: in fact, the coherent state $\psi^-(z_1,z_2)$, due to the terms $\sqrt{n!}$ in the denominator, has a higher contribution coming from the vectors $\tilde\psi^{-}_n=\psi^{+}_n$ when $n<V^2$, so that the "+" vectors contribute predominantly in the construction of both the coherent states $\varphi^+(z_1,z_2)$ and $\psi^-(z_1,z_2)$ when $V$ is quite large. They exhibit a gain in the first/second component compared to the second/first, respectively, and overall, a shifted concentration peak compared to the $V=0$ scenario. Similar patterns, though not depicted here, are observed for the other coherent states $\varphi^-(z_1,z_2)$ and $\psi^+(z_1,z_2)$. Comparable results are seen for the second class of coherent states $\eta^+(z_1,z_2)$ and $\xi^-(z_1,z_2)$, as shown in Figures \ref{fig7}-\ref{fig8}. Here, the first component of $\eta^+(z_1,z_2)$ experiences a gain compared to its second component, and similarly, the second component of $\xi^-(z_1,z_2)$ shows a gain compared to its first. This illustrates how the extensive broken phase tends to accentuate this gain and loss effect. We notice that this behavior is highlighted only for a sufficiently large value of $V$, whereas for a low value of $V$ this phenomenon is not particularly evident.
 
%

\section{Conclusions}\label{sect4}

In this paper we have considered a self-adjoint Hamiltonian for graphene, and its non self-adjoint version where a chemical potential is also included. In the first case we have shown how the lack of a ground state of the Hamiltonian suggests the introduction of ladder operators of a special kind, and we have proposed a related class of coherent states. These are better defined on two separated subspaces of the Hilbert space of the full system. 

A similar functional structure is recovered also in presence of a chemical potential, but with a {\em serious} difference: complex eigenvalues appear, which are connected with the fact that the Hamiltonian is no longer self-adjoint, and coherent states must be replaced by bicoherent states. In particular, after proposing a first {\em almost standard} class of these vectors, we introduce a different class, which is more closely related to the Hamiltonian of the system, since these other bicoherent states are eigenvectors of some special lowering operators which can be used to factorize the Hamiltonian of the system, in presence of the chemical potential. In this context we have examined the role of $V$ and its impact as its magnitude increases, demonstrating that a higher strength leads to a {\em larger} broken $\mathcal{P}\mathcal{T}$ phase characterized by the emergence of gain and loss phenomena. Specifically, when $V$ increases significantly, the non-Hermitian elements (associated with gain and loss) within the Hamiltonian begin to predominate over the kinetic components. Consequently, the system's behavior is primarily influenced by processes of amplification and attenuation, rather than by the conventional dynamics driven by the kinetic energy in presence of a magnetic field. The role of squeezed states in the context discussed here has still to be understood. This is part of our future plans.

\bibliographystyle{unsrt}

\end{document}